\documentclass[acmsmall]{acmart}
\AtBeginDocument{%
  }

\begin{document}

\title{Descriptor-Based Object-Aware Memory Systems: A Comprehensive Review}

\author{Dong Tong}
\email{tongdong@pku.edu.cn}
\orcid{0000-0003-0987-6351}
\affiliation{%
  \institution{School of Computer Science, Peking Univeristy}
  \city{Beijing}
  \state{Beijing}
  \country{China}
}

\renewcommand{\shortauthors}{Dong Tong}

\begin{abstract}
  The security and efficiency of modern computing systems are fundamentally undermined by the absence of a native architectural mechanism to propagate high-level program semantics, such as object identifier, bounds, and lifetime, across the hardware/software interface. This paper presents a comprehensive survey of the architectural paradigm designed to bridge this semantic gap: descriptor-based, object-aware memory systems. By elevating the descriptor to a first-class architectural abstraction, this paradigm enables hardware to dynamically acquire and enforce the rich semantics of software-defined objects.
  
  This survey systematically charts the evolution and current landscape of this approach. We establish the foundational concepts of memory objects and descriptors and introduce a novel taxonomy of descriptor addressing modes, providing a structured framework for analyzing and comparing diverse implementations. Our unified analysis reveals how this paradigm holistically addresses the intertwined challenges of memory protection, management, and processing. As a culminating case study, we re-examine the CentroID model, demonstrating how its hybrid tagged-pointer encoding and descriptor processing mechanisms embody the path toward practical and efficient object-aware designs. Finally, we outline how the explicit cross-layer communication of object semantics provides a foundational research direction for next-generation cache hierarchies, unified virtual memory, and even 128-bit architectures.
\end{abstract}




\maketitle

\section{Introduction}
The rise of Artificial Intelligence (AI) and the Internet of Everything (IoE) has made data-intensive computing and cybersecurity paramount, driving the innovations on next-generation architectures. While Domain-Specific Architectures (DSAs) have been widely explored to meet these computational requirements, their associated memory systems persist as a critical bottleneck \cite{gholamiAIMemoryWall2024,hennessyNewGoldenAge2019,hill21stCenturyComputer2016,jagadishBigDataIts2014,reedExascaleComputingBig2015}. 
These traditional memory systems, which rely on virtual memory and cache hierarchies based on fixed-size blocks, exhibit significant limitations in security, reliability, performance, and energy efficiency
\cite{denningVirtualMemory1970,hennessyComputerArchitectureQuantitative2019,jacobVirtualMemoryContemporary1998}.
Specifically, they inherently lack native and holistic support for fine-grained memory safety \cite{szekeresSoKEternalWar2013,witchelMondrianMemoryProtection2002,woodruffCHERICapabilityModel2014} and efficient management of ultra-large-scale data objects \cite{basuEfficientVirtualMemory2013,gonzalezProfilingHyperscaleBig2023,kanevProfilingWarehousescaleComputer2015}. 
Addressing these systemic limitations fundamentally demands architectural breakthroughs across three intertwined fronts: memory protection, memory management, and memory-centric computing.

\begin{figure}[t]
	\centering
	\includegraphics[width=\linewidth]{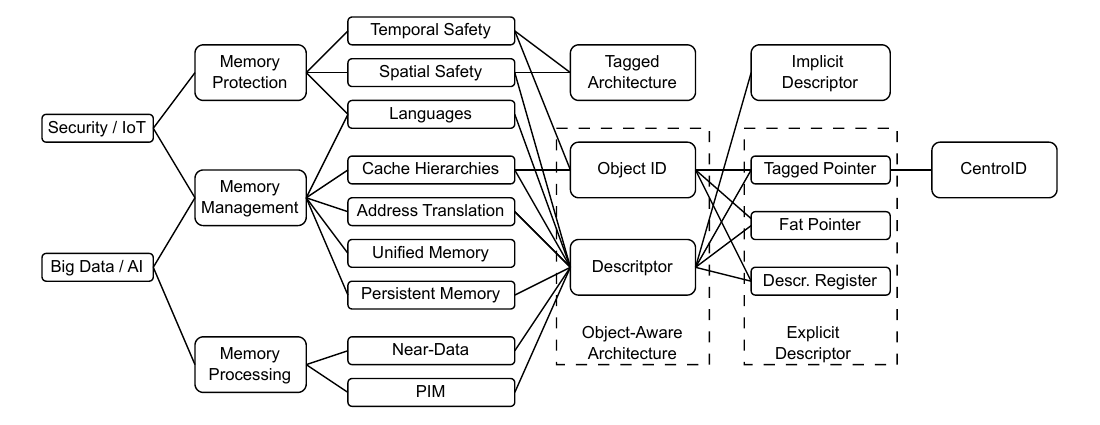}
	\caption{The relationship of concepts in this survey}
	\Description{outline of this paper}
\end{figure}

\textbf{Memory Protection} is paramount for cybersecurity \cite{hennessyNewGoldenAge2019,songSoKSanitizingSecurity2019,szekeresSoKEternalWar2013}. 
Empirical data consistently confirms that memory safety violations account for approximately 70\% of all critical vulnerabilities, serving as the primary entry point and fundamental root cause for the majority of successful exploits 
\cite{millerTrendsChallengesStrategic2019}. 
Memory safety vulnerabilities are often simplistically attributed to the pervasive use of unsafe languages like C/C++. This has spurred the development of numerous software-based exploit mitigation techniques and vulnerability detection tools 
\cite{songSoKSanitizingSecurity2019,szekeresSoKEternalWar2013}. 
Recently, memory-safe languages have been widely employed in new software development. However, even languages like Rust, which uses formal methods to eliminate temporal memory errors, still suffer from critical limitations. They incur non-negligible overhead from runtime checks 
\cite{zhangUnderstandingRuntimePerformance2022} 
and, more recently, have been shown to be vulnerable to an emerging class of microarchitectural side-channel attacks 
\cite{canellaSystematicEvaluationTransient2019,kocherSpectreAttacksExploiting2020}.

The root cause of memory safety vulnerabilities stems from the lack of a pointer abstraction within mainstream Instruction Set Architecture (ISA)
\cite{carterHardwareSupportFast1994,deviettiHardboundArchitecturalSupport2008,dhawanArchitecturalSupportSoftwareDefined2015,kwonLowfatPointersCompact2013,woodruffCHERICapabilityModel2014}, 
a deficiency manifested in three critical aspects: the absence of a dedicated pointer type causes address calculation to be substituted with integer arithmetic; instructions involving memory addresses calculation cannot retrieve the high-level semantics, such as bounds and lifetime, of corresponding memory objects; and the processor hardware provides no built-in, atomic safety checking for address calculation results. 
Although pioneering projects such as CHERI 
\cite{amarCHERIoTCompleteMemory2023,grisenthwaiteArmMorelloEvaluation2023,watsonCHERIHybridCapabilitySystem2015,woodruffCHERICapabilityModel2014} 
have made substantial progress by introducing capability-based pointers, their reliance on fat- pointer encoding introduces systemic challenges, most notably persistent compatibility issues with traditional Application Binary Interfaces (ABIs) that still await a definitive resolution.

\textbf{Memory management} systems for big data processing fundamentally rely on virtual memory to enable physical storage sharing and coarse-grained process isolation. These systems further leverage multi-level cache hierarchies and Non-Uniform Memory Access (NUMA) architectures to place hot data near processing units 
\cite{hennessyComputerArchitectureQuantitative2019}. 
As the working sets of modern big data applications expand relentlessly, the demand for main memory capacity in computing platforms grows correspondingly. In response, emerging technologies such as Non-Volatile Memory (NVM) 
\cite{coburnNVHeapsMakingPersistent2011,volosMnemosyneLightweightPersistent2011,zhangG10EnablingEfficient2023} and the Compute Express Link (CXL) 
\cite{dassharmaIntroductionComputeExpress2024,liPondCXLBasedMemory2023,marufTPPTransparentPage2023}  interconnect interface are driving a paradigm shift in memory architecture—toward heterogeneous, disaggregated, tiered, and unified virtual memory systems 
\cite{dulloorDataTieringHeterogeneous2016,limDisaggregatedMemoryExpansion2009,meswaniHeterogeneousMemoryArchitectures2015}.

Traditional virtual memory systems, which rely on multi-level radix tree page tables 
\cite{denningVirtualMemory1970,jacobVirtualMemoryContemporary1998}, 
incur substantial memory overhead and performance degradation due to the cost of traversing these multi-level structures. In response, advanced memory management schemes have converged on two variable-length, object-aware strategies: either allocating a contiguous physical address space for an object's contiguous virtual address range 
\cite{basuEfficientVirtualMemory2013,chenFlexPointerFastAddress2023,karakostasRedundantMemoryMappings2015,zhaoContiguitasPursuitPhysical2023},
or maintaining independent linear page tables for individual large objects \cite{zhangDirectMemoryTranslation2024}. Both approaches significantly reduce translation and storage overhead. Furthermore, propagating object-granularity semantic information—such as object identifiers and sizes—across the cache hierarchy can improve the learning of memory access patterns. This, in turn, enhances cache management policies such as prefetching and reduces average access latency throughout the memory hierarchy.

\textbf{Memory-centric computing} and near-data processing architectures have emerged for overcoming energy efficiency bottlenecks 
\cite{ahnScalableProcessinginmemoryAccelerator2015,ahnPIMenabledInstructionsLowoverhead2015,boroumandGoogleWorkloadsConsumer2018,mutluMemoryCentricComputingSolving2025}. 
These architectures aim to offload memory-intensive operations, such as data movement, to the data's proximity or directly within the memory modules, thereby avoiding low computational-density data transfer between the memory and processing units. However, contemporary implementations typically rely on the use of configuration registers and trigger instruction, resulting in poor programmability and limited flexibility. To fully realize the potential of these architectures, intrinsic support for memory object-granularity operations must be integrated at the instruction set level. This integration enables applications to more effectively control memory access and offloaded computation processes, drastically improving system utilization and performance.

Based on the preceding analysis, the intertwined challenges of memory safety, memory management, and memory processing collectively reveal a fundamental architectural mismatch: contemporary memory architectures operate on fixed-size address blocks as their basic element, whereas software manipulates variable-length data objects with rich semantics. Furthermore, there exists no effective mechanism to propagate these high-level semantic insights to the underlying hardware, creating a critical semantic gap that lies at the root of current system inefficiencies and vulnerabilities.

\textbf{Object-Aware Memory Systems} can offer a unified architectural approach to simultaneously resolve the challenges detailed previously. An object-aware memory system fundamentally shifts the hardware's operating view from a uniform array of address blocks to logical, variable-sized objects explicitly defined by their metadata, such as identifiers, boundaries, lifetimes, and access permissions 
\cite{levyCapabilitybasedComputerSystems1984,rattnerObjectbasedComputerArchitecture1980,wangOASISObjectAwarePage2025,welchInvestigationDescriptorOriented1976}. 
By operating directly on these object-level semantics, the architecture is empowered to enforce fine-grained security policies and transparently optimize data placement and movement.

The design of an effective object-aware memory system must fulfill three interdependent architectural mandates:

\textbf{Comprehensive Support.} The architecture must comprehensively represent all memory objects across the entire memory hierarchy, while fully supporting the functional requirements of protection, management, and memory-centric computing.

\textbf{Scalability and Efficiency.} The system must scale efficiently across diverse computing environments—from embedded devices to data centers—maintaining low runtime overhead and minimal complexity in both hardware and software.

\textbf{Backward Compatibility.} The design must ensure seamless interoperability with the existing software ecosystem, particularly by preserving full Application Binary Interface (ABI) compatibility to enable immediate adoption and reduce porting effort.

Hardware support for variable-length memory objects is primarily realized through two distinct paradigms: the location-based and the identifier-based approaches  
\cite{nagarakatteWatchdogHardwareSafe2012,sasakiPracticalByteGranularMemory2019,songSoKSanitizingSecurity2019}. 
Unlike the location-based method, which attaches handful metadata directly to each memory unit 
\cite{dhawanArchitecturalSupportSoftwareDefined2015,feustelAdvantagesTaggedArchitecture1973,witchelMondrianMemoryProtection2002}, 
the identifier-based approach is architected around a unique, centralized descriptor for each object 
\cite{daleyVirtualMemoryProcesses1968,dennisSegmentationDesignMultiprogrammed1965,fabryCapabilitybasedAddressing1974,grahamProtectionInformationProcessing1968,levyCapabilitybasedComputerSystems1984}. 
This descriptor-centric model is demonstrably superior in comprehensively fulfilling the three aforementioned design mandates. The descriptor is required to encapsulate the object's core attributes, such as its address range or a unique identifier, and can be extended to involve rich metadata, including access permissions, address translation details, high-level semantic features, and other critical attributes.

A paramount requirement for descriptor-based architectures is the ability to efficiently and accurately retrieve or generate the corresponding object descriptor for any instruction executing an address calculation. However, contemporary research efforts encounter two significant, systemic hurdles. First, architectural and hardware resource constraints limit the total number of available descriptors 
\cite{carterHardwareSupportFast1994,saltzerProtectionControlInformation1974}. 
Second, the complex descriptor data structures required to manage variable-length objects introduce significant query latency. These limitations directly affect the practical viability and efficiency of descriptor-based memory systems 
\cite{chenMetaTableLiteEfficientMetadata2021,deviettiHardboundArchitecturalSupport2008,fabryCapabilitybasedAddressing1974,kimHardwarebasedAlwaysOnHeap2020,oleksenkoIntelMPXExplained2018,sharifiCHEx86ContextSensitiveEnforcement2020}.

Although its origins date back to the time-sharing systems of the 1960s 
\cite{dennisProgrammingSemanticsMultiprogrammed1966,glaserSystemDesignComputer1965,levyCapabilitybasedComputerSystems1984,mayerArchitectureBurroughsB50001982}, 
the object-aware memory paradigm has recently experienced a significant architectural resurgence. This renewal is driven by the confluence of stringent security mandates 
\cite{dhawanArchitecturalSupportSoftwareDefined2015,watsonCHERIHybridCapabilitySystem2015,witchelMondrianMemoryProtection2002,woodruffCHERICapabilityModel2014} 
and pressing efficiency demands in the modern data-intensive computing landscape 
\cite{basuEfficientVirtualMemory2013,chenFlexPointerFastAddress2023,guptaRebootingVirtualMemory2021,karakostasRedundantMemoryMappings2015,tsaiRethinkingMemoryHierarchy2018,wangOASISObjectAwarePage2025,zhangDirectMemoryTranslation2024}.

\vspace{1\baselineskip}

This survey presents a systematic and comprehensive review of object-aware memory architectures, centering on the object descriptor as the core abstraction. Figure 1 illustrates the foundational concepts and their interrelationships discussed herein. The paper proceeds as follows:

Section 2 introduces the fundamental concepts, including memory objects, descriptors, and the imperative for address authentication. Section 3 provides a historical analysis of object-aware architectures and their applications. Section 4 proposes a taxonomy of descriptor addressing modes and their implementation mechanisms. Subsequently, Section 5 revisits the CentroID-based hybrid tagged pointer scheme and its architectural implications, demonstrating its capability to support a complete and extensible object-aware system while maintaining pointer width ABI compatibility. Finally, Section 6 discusses relevant fields that can leverage object-aware approaches. It is important to note that this survey is not exhaustive regarding related work in each area, but rather focuses on providing a limited number of canonical references for illustrative purposes. 

\section{Object-Aware Memory System}
\subsection{Memory Object}
An object is fundamentally defined as a region of storage in the execution environment that possesses a unique identifier and whose stored content represents a specific data value. Drawing from the C programming model 
\cite{kernighanProgrammingLanguage2014}, 
an object's interpretation is primarily governed by its data type and its lifetime. The object's logical size, explicitly determined by program semantics, gives it a definite address range, typically expressed as the tuple <name, length> (identifier and byte-grained size). Internally, program locate individual components using <name, offset> 
\cite{dennisSegmentationDesignMultiprogrammed1965}.

An object's memory space is allocated either statically (by the compiler/linker for stack and global objects) or dynamically (via heap management, e.g., 
malloc \cite{kernighanProgrammingLanguage2014}, 
or OS interfaces for large objects, e.g., 
mmap \cite{bovetUnderstandingLinuxKernel2008}). 
Since most high-performance memory allocators use a binning strategy 
\cite{leijenMimallocFreeList2019,ibnziadNoFATArchitecturalSupport2021}, 
the physical allocation size is frequently rounded up to the nearest available size class. Consequently, the actual address range of the allocated storage is represented by the notation (Base, Bound), which denote the first and the last address of the allocated object. Henceforth, unless otherwise specified, the term "object" refers specifically to the allocated memory object.

\subsection{Characteristics of Memory Objects}
The dynamic statistical properties of objects in contemporary programs offer critical insights for optimizing memory system design. Research has illuminated several key characteristics:

\textbf{Object Volume:} The maximum number of live objects per application frequently reaches 
millions \cite{bodikABCDEliminatingArray2000,jiUnderstandingObjectlevelMemory2017}.

\textbf{Size Distribution and Footprint}: Objects exhibit a bimodal size distribution. The vast majority (up to 99\%) are small-sized (< 1 KB) but account for a minority of total memory consumption. Conversely, a small fraction of large objects dominates the overall memory footprint. This pattern holds across diverse computational domains, including 
scientific/desktop computing \cite{basuEfficientVirtualMemory2013,chenFlexPointerFastAddress2023,jiUnderstandingObjectlevelMemory2017}, 
warehouse environments \cite{gonzalezProfilingHyperscaleBig2023,kanevProfilingWarehousescaleComputer2015,zhouCharacterizingMemoryAllocator2024}, and 
serverless architectures \cite{wangMementoArchitecturalSupport2023}.

\textbf{Lifetime and Access Density: }Object lifetimes also exhibit a bimodal distribution, with most small objects being short-lived, while large objects (especially in scientific applications) tend to be long-lived 
\cite{jiUnderstandingObjectlevelMemory2017,wangMementoArchitecturalSupport2023}. 
Furthermore, large objects often exhibit high access density and strong spatial locality 
\cite{jiUnderstandingObjectlevelMemory2017}.

\textbf{Access Sparsity:} During any single execution or within specific time intervals, only a limited portion of large data objects tends to be actively accessed 
\cite{linDrGPUMGuidingMemory2023,zhangG10EnablingEfficient2023}.

In summary, these observations confirm the cross-domain prevalence of object heterogeneity, underscoring the imperative for memory system designs that employ differentiated management strategies.

\subsection{Descriptors}
In general terms, a descriptor is a stored information that describes how other information is organized and accessed. In the context of computer architecture, a descriptor serves as the core data structure defining a memory object. It is a protected storage value that ether holds or points to the object’s description 
\cite{saltzerProtectionInformationComputer1975}. 

\begin{figure}[t]
	\centering
	\includegraphics[width=\linewidth]{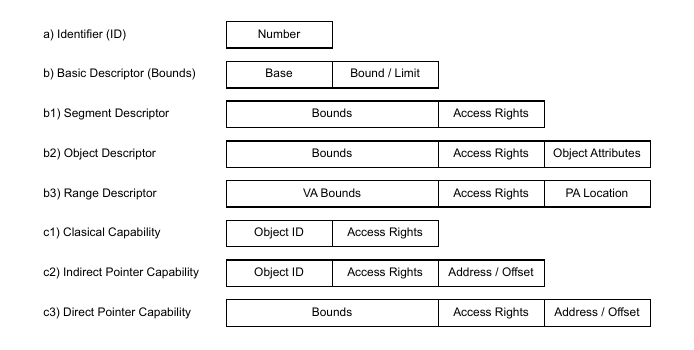}
	\caption{Formats of the descriptor and capability}
	\Description{descriptor formats}
\end{figure}

A descriptor can function as a unique object identifier (ID) for a memory object (Figure 2-a), facilitating the indexing and retrieval of the full object descriptor.  Furthermore, the memory system can leverage the object ID to infer runtime memory access behavior, which in turn supports advanced management decisions such as prefetching and data placement.

A basic descriptor is the bounds descriptor (Figure 2-b), often simply calls 'Bounds', which includes the address range of the memory object. This is typically represented as (Base, Limit) or (Base, Bound), where Limit indicates the size of the object in the storage units. The basic descriptor is essential for bounds checking, which determine whether a memory access address falls within the legal range of the object. Depending on application requirements, this basic descriptor can be extended to incorporate rich metadata such as access permission bits, virtual-to-physical address translation information, and high-level program semantic tags. This structured mechanism of metadata encapsulation provided by descriptors enables fine-grained memory protection and management capabilities in modern computing systems.

Building upon the concepts of the object ID and the basic descriptor, specific descriptor types can be derived based on their functional scope. A segment descriptor (Figure 2-b1) inherits from a basic descriptor but adds access rights for memory protection and isolation 
\cite{childsProcessorFamilyPersonal1984,daleyVirtualMemoryProcesses1968,dennisSegmentationDesignMultiprogrammed1965,grahamProtectionInformationProcessing1968,levyCapabilitybasedComputerSystems1984}. 
An object descriptor (Figure 2-b2), in turn, inherits from the segment descriptor and incorporates object attributes to facilitate cross-layer semantic propagation 
\cite{colwellPerformanceEffectsArchitectural1988,dallyObjectOrientedArchitecture1985,levyCapabilitybasedComputerSystems1984}. 
Finally, a range descriptor (Figure 2-b3) also builds upon the segment descriptor by including physical address ranges, thereby exploiting in-object space continuity to mitigate the performance penalty of page table walks 
\cite{chenFlexPointerFastAddress2023,karakostasRedundantMemoryMappings2015}. 
A more detailed discussion of these descriptor formats, along with arguments about capability addressing (Figure 2-c1, 2-c2, 2-c3), is provided in the section 3 and 4.

\subsection{Memory Protection and Address Authentication}
The Object-Aware Memory System represents an ancient yet novel computational paradigm fundamentally organized around the characteristics of the objects in program. It employs a hardware/software codesign approach, enabling the hardware to dynamically acquire and process key object attributes, including object identifiers, address ranges, access permissions, and other high-level semantics. Ultimately, it bridges the inherent "semantic gap" present in conventional systems, thereby facilitating the comprehensive memory protection, flexible memory processing and optimized memory management strategies.

Memory protection is a cornerstone mechanism for ensuring computer system security and reliability, and thus forms a critical foundation for the object-aware memory systems. Only when memory safety is guaranteed can the memory management and processing hardware accurately acquire high-level semantic information. Classical security theory typically mandates a two-phase process of authentication and authorization, where an access matrix defines permissible operations at the granularity of individual memory objects 
\cite{grahamProtectionPrinciplesPractice1971,lampsonProtection1971,saltzerProtectionControlInformation1974,saltzerProtectionInformationComputer1975}.

The fixed-size granularity of conventional memory access (e.g., bytes or words) creates a fundamental misalignment with the object-grained access control required for complete memory protection. To establish this safety, memory addresses must be authenticated to ensure they fall strictly within the boundaries of the intended object, thereby preventing any out-of-bounds address arithmetic or illegal access. Contemporary mainstream architectures rely on software-based bounds checking, a method that introduces significant performance overhead and inherent security risks. Therefore, object-aware architectures critically need to implement hardware-enforced bounds-checking mechanisms. This hardware-level enforcement is essential to not only guarantee security but also to ensure consistent object semantics across all system layers.

Classical approaches to memory safety protection can be broadly categorized into three distinct methods: page-based, tag-based, and descriptor-based schemes. The first two fall under the category of location-based schemes, where additional metadata (such as permissions or tags) is locally associated with each memory location (page or word). In contrast, the descriptor-based approach represents an identifier-based memory protection scheme.

Figure 3-a illustrates the critical importance of address authentication using a concrete example. In this scenario, the subject of the access matrix is a memory-access instruction, and the objects are the individual bytes of three adjacent memory array in the same page: Object A, Object B, and Object C. The instruction sequentially generates three byte-access operations targeting addresses A+2, A+5, and A+8.

\textbf{Page-based memory protection} inherently lacks object-grained bounds checking. Since Objects A, B, and C reside within the same memory page, where all bytes share identical permissions P, then the permission granted to access A+2 implicitly grants access to A+5 and A+8, even if these accesses fall outside the bounds of Object A, resulting in a severe memory safety violation. Consequently, all three addresses, A+2, A+5, and A+8, may be issued directly to the underlying hardware memory subsystem protection unit. Furthermore, these addresses may be issued directly to the underlying hardware memory subsystem even during speculative execution, bypassing any supplementary software checks. 

\textbf{Tag-based memory protection}’s security checks occur after the memory access. The tag associated with the accessed data is read and compared against the tag associated with the instruction or the pointer. As memory tags typically comprise a limited number of bits, adjacent objects can often be assigned distinct tags, while non-adjacent objects may still share the same tag. In our running example, assuming Object A and Object C share tag T, while Object B possesses a different tag, access to A+2 would also implicitly permit access to A+8; however, access to A+5 would be correctly blocked. This demonstrates that while tag-based protection offers finer granularity than its page-based counterpart, it still does not constitute a complete memory safety solution because of tag collision and the lack of precise bounds information.

\textbf{Descriptor-based memory protection} is designed to query or dynamically generate object boundaries from a given pointer and perform a mandatory bounds-checking on the address calculation result. If an out-of-bounds access is detected, the memory request is blocked and not issued to the memory subsystem. As Figure 3-a illustrates, an address authentication mechanism implemented at the address generation stage would successfully intercept and disallow the memory demands for A+5 and A+8. Consequently, this approach can achieve complete spatial memory safety, preventing the violations at A+5 and A+8 even if the access matrix grants the same permission (D) to all three adjacent objects.

\begin{figure}[t]
	\centering
	\includegraphics[width=\linewidth]{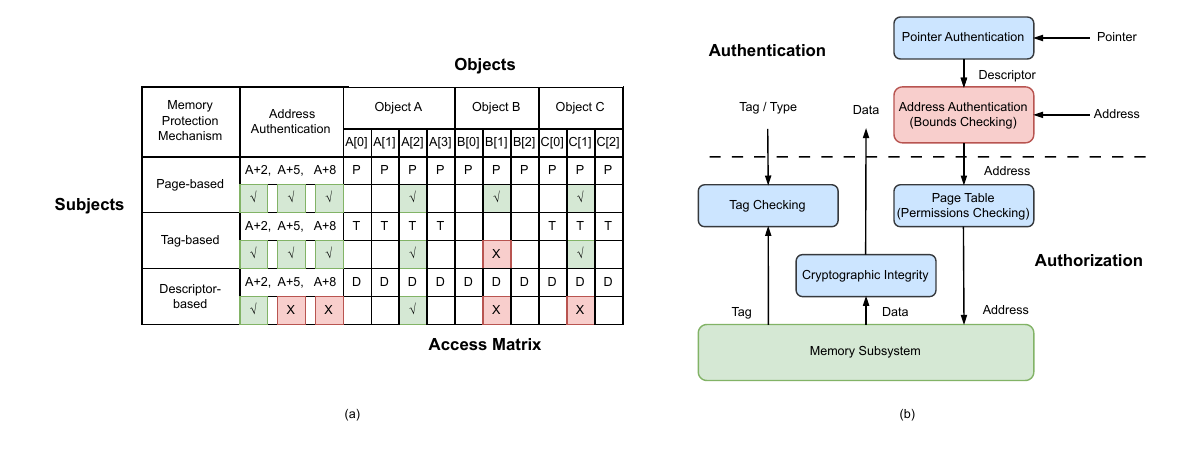}
	\caption{Object-Aware Memory Protection. (a) Access Martix (b) Security Checking Phases}
	\Description{Address Authentication}
\end{figure}

\vspace{1\baselineskip}

Figure 3-b outlines a multi-stage security check mechanism that a load instruction might employ to achieve end-to-end safety. This process ensures protection across the execution pipeline: 

\textbf{Pointer authentication} check the legitimacy of the instruction's source pointer operand \cite{liljestrandPACItPointer2019}. This crucial step ensures the correct retrieval of the target memory object's descriptor.

\textbf{Address authentication} performs a hardware-enforced mandatory bounds checking to confirm that the calculated address falls within the valid range specified by the descriptor. 

\textbf{Access Control} checks that if the requested operation complies with the permissions and privilege levels defined in the corresponding Page Table Entry (PTE).

\textbf{Data Integrity Verification} could be performed to detect any tampering by malicious software or physical attacks \cite{lemayCryptographicCapabilityComputing2021}. 

\textbf{Tag Checking} can be performed if a tagged memory mechanism is utilized. The final check at the retirement stage compares the data tag against the pointer tag or instruction type to ensure semantic integrity. 
\cite{feustelAdvantagesTaggedArchitecture1973,jeroTAGTaggedArchitecture2023,sullivanImplicitMemoryTagging2023}.

It is crucial to note that while these last two methods provide valuable security functions, their post-access nature means they are inherently unable to prevent microarchitectural side-channel attacks that occur during the access phase.

\section{History}
\subsection{Overview}
Object-aware memory systems typically fall into two fundamental categories: tag-based and descriptor-based schemes. Tag-based mechanisms define objects by assigning different tags to memory locations. This approach benefits from simpler hardware extension and better compatibility  
\cite{dhawanArchitecturalSupportSoftwareDefined2015,feustelAdvantagesTaggedArchitecture1973,jeroTAGTaggedArchitecture2023,joannouEfficientTaggedMemory2017,sasakiPracticalByteGranularMemory2019,sinhaPracticalMemorySafety2018,sullivanImplicitMemoryTagging2023,tiwariSmallCacheLarge2008,woodruffCHERICapabilityModel2014}. 
However, it suffers from two critical drawbacks: first, it fundamentally cannot precisely delineate the exact address boundaries of an object. Second, the limited bit-width of memory tags severely restricts the system's capacity to support a diverse range of object types. Consequently, tag-based methods are best suited for coarse-grained protection, often acting as a 
'blacklist' \cite{sasakiPracticalByteGranularMemory2019}, 
or simple architectural extensions (such as pointer integrity), proving ultimately insufficient for the comprehensive protection and management demands of a sophisticated object-aware computing architecture.

In contrast, the descriptor-based object-aware architecture is proposed to transfer high-level semantic information, such as precise object address ranges or ID, alongside memory addresses to the underlying hardware 
\cite{fabryCapabilitybasedAddressing1974,rattnerObjectbasedComputerArchitecture1980,welchInvestigationDescriptorOriented1976}. 
This crucial capability facilitates fine-grained, 'whitelist'-style memory protection 
\cite{sasakiPracticalByteGranularMemory2019} 
and enables efficient memory management and processing for large-scale data objects. Owing to its precision in memory boundary description and extensible support for other metadata, this architectural paradigm has emerged as the predominant technical path in the computer architecture field. Consequently, this paper employs a unified taxonomy, classifying all architectures that feature hardware-level support for object descriptors or ID as descriptor-based object-aware memory architectures, and systematically traces their chronological evolution. Figure 4 illustrate a brief timeline of previous works.

\subsection{Two-dimension address space architecture era}
During the 1950s and 1960s, pioneering implementations of memory protection for time-sharing systems relied on processors utilizing dedicated descriptor registers as the explicit or implicit operand. These registers held the basic descriptors (Figure 2-b). Hardware circuits then performed mandatory bounds checking on every memory access. This pioneering mechanism was evident in historical machines such as the 
IBM Stretch \cite{coddMultiprogrammingSTRETCHFeasibility1959}, 
Burroughs B5000 \cite{mayerArchitectureBurroughsB50001982}, and the 
CDC 6600 \cite{thorntonParallelOperationControl1964}. 
Notably, its core design philosophy continues to be reflected in contemporary processors, exemplified by the explicit bound registers found in Intel's Memory Protection Extensions 
(MPX) \cite{oleksenkoIntelMPXExplained2018}.

The two-dimensional address space architecture fundamentally couples logical identifiers with the memory addresses by assigning a dedicated descriptor to each program object. These descriptors are logically managed within a descriptor table, allowing for efficient, direct retrieval via unique object identifiers. This mechanism can naturally map programming-level objects to memory segments, where an object's symbolic name directly corresponds to its unique object ID, translating program-level <name, offset> references into architectural <ID, offset> addresses. Historically, such architectures were often employed to resolve the addressing limitations imposed by narrow machine word lengths (e.g., 16 or 24 bits) in early computer designs. For high-performance implementation, a dedicated descriptor cache can be integrated at the hardware level, utilizing the object ID as a tag to significantly reduce lookup latency and enhance overall system efficiency.

\textbf{Segmented Architecture (1960s – 1970s)}

The segmented memory architecture simultaneously provides both a two-dimensional addressing scheme and a dedicated memory protection system simultaneously. This was essential for enabling dynamic memory allocation, position-independent code, inter-process data sharing, and resource isolation 
\cite{dennisSegmentationDesignMultiprogrammed1965}. 
The memory protection relies on segment descriptors to control access permissions and boundary limits. Prominent historical examples are the 
Multics project \cite{daleyVirtualMemoryProcesses1968,saltzerProtectionControlInformation1974} and 
the Intel 80x86 architecture \cite{childsProcessorFamilyPersonal1984}, 
where segmentation organizes the virtual address space and enforces protection, often in conjunction with paging for managing physical memory 
\cite{mutluMemoryCentricComputingSolving2025}. 
However, in multi-programming environments, enabling shared data access often need frequent Operating System intervention to maintain proper isolation and manage complex copying mechanisms.

\textbf{Capability-Based Architecture (1970s -1980s)}

The capability architecture provides an efficient and secure solution for systems requiring frequent switching of protection domains 
\cite{dennisProgrammingSemanticsMultiprogrammed1966,fabryCapabilitybasedAddressing1974,levyCapabilitybasedComputerSystems1984}. 
Its core principle is the granting of object access rights through unforgeable security tokens known as capabilities. Early implementations built upon the segment descriptor table to realize the capability, which contains a unique ID and its associated access rights (Figure 2-c1). System support for efficient capability transfer across protection boundaries was achieved through dedicated capability registers and specialized hardware instructions. Furthermore, it facilitates permission degradation to implement fine-grained, differentiated access control to the same object, and ensures the integrity and unforgeability of the capability itself via memory tagging technology. This design paradigm has been successfully validated in classic systems such as the 
IBM System/38 \cite{berstisSecurityProtectionData1980} and the 
Cambridge CAP Computer \cite{needhamCambridgeCAPComputer1977}.

\textbf{Object-Based Architecture (1980s -1990s)}

An object-based architecture 
\cite{ishikawaDesignObjectOriented1984,rattnerObjectbasedComputerArchitecture1980} 
has the ability to raise the level of the hardware/software interface and integrate the concepts of data abstraction, domain-based protection and high-level system functionality. The crucial impetus for object-based architecture was the research and development of object-oriented programming language. While early object-oriented architectures often leveraged tag-based systems in conjunction with microprogrammed control, the more advanced capability-based architectures provide true hardware-level object-oriented support. This was achieved by embedding critical metadata (e.g., type information) directly within the capability descriptors (Figure 2-c2) and augmenting the instruction set with object-granularity operations. A landmark examples of this design philosophy include the Intel 432 
\cite{colwellPerformanceEffectsArchitectural1988}.

\begin{figure}[t]
	\centering
	\includegraphics[width=\linewidth]{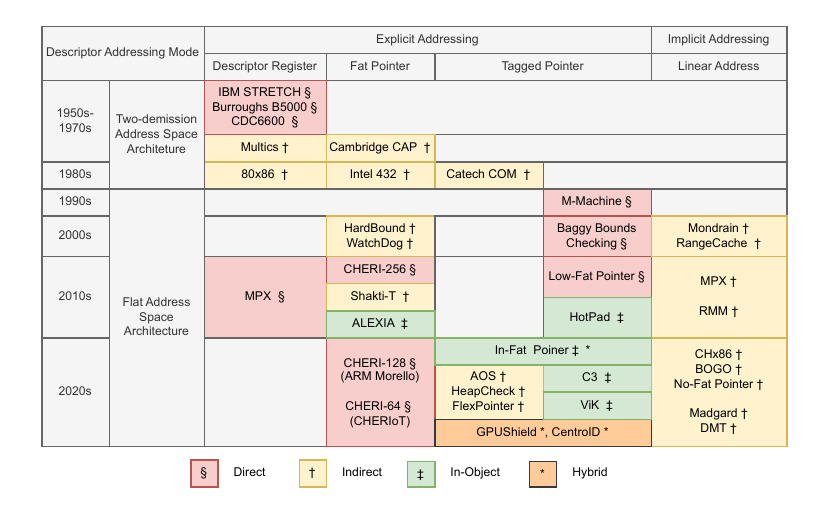}
	\caption{A Brief History of the Descriptor-based Object-Aware systems}
	\Description{History}
\end{figure}

These early segment-based object-aware systems faced two critical limitations: 1) the finite number of segment ID was insufficient to manage the massive number of memory objects. 2) the multi-level table lookups resulted in substantial memory access latency 
\cite{carterHardwareSupportFast1994,saltzerProtectionControlInformation1974}. 
To this end, historical solutions were proposed: the 
IBM System/38 \cite{berstisSecurityProtectionData1980} 
employed a fixed-length memory segment design to reduce descriptor management and implementation complexity. 
Floating Address scheme \cite{dallyObjectOrientedArchitecture1985} 
was proposed to resolve the conflict between small data objects and the need for a large ID space. This method encodes object size information within the address itself, using a floating-point-like scheme to simultaneously compress the offset bit-width and expand the segment identifier count. These historical efforts underscore the persistent trade-offs between address space size, descriptor count, and lookup efficiency, paving the way for modern solutions.

\subsection{Flat address space architecture era}
With the evolution of hardware and programming technologies, particularly the rise of 
the RISC architecture \cite{pattersonCaseReducedInstruction1980}, 
the flat address space has emerged as the dominant architectural paradigm 
\cite{carterHardwareSupportFast1994,koldingerArchitectureSupportSingle1992,witchelMondrianMemoryProtection2002}. 
Its prevalence is driven by its benefits in simplified hardware and software implementation, superior cost-effectiveness, and enhanced flexibility and portability, enabling high performance with significantly reduced hardware complexity.

However, the flat address space architecture inherently lacks cross-layer high-level semantic mechanisms and build-in access control mechanism from segment descriptors. This deficiency has led to critical bottlenecks and resource contention issues in memory safety and memory management. Consequently, substantial research has been devoted to descriptor-based object-aware memory systems to overcome these limitations.

\subsubsection{Memory Protection}

The pursuit of hardware-assisted memory safety for flat address space architecture began with the M-Machine's 
Guarded Pointer \cite{carterHardwareSupportFast1994} 
in the 1990s, pioneering a descriptor generation approach. The 2000s saw the architectural focus shift toward hardware-assisted vulnerability and exploits detection, primarily through 
Tagged Architectures \cite{tiwariSmallCacheLarge2008,witchelMondrianMemoryProtection2002} 
and dynamic techniques like 
DIFT \cite{suhSecureProgramExecution2004}, 
PUMP \cite{dhawanArchitecturalSupportSoftwareDefined2015}, 
ASAN \cite{serebryanyAddressSanitizerFastAddress2012}, and 
HDFI \cite{songHDFIHardwareAssistedDataFlow2016}. 
These efforts ultimately fell short of providing complete memory safety guarantees. 

The 2010s marked a pivot toward descriptor-based hardware-enforced security 
\cite{akritidisBaggyBoundsChecking2009,deviettiHardboundArchitecturalSupport2008}, 
focusing on both 
spatial \cite{krishnakumarALEXIAProcessorLightweight2019,kwonLowfatPointersCompact2013,menonShaktiTRISCVProcessor2017,oleksenkoIntelMPXExplained2018,woodruffCHERICapabilityModel2014} and 
temporal memory safety \cite{choViKPracticalMitigation2022,nagarakatteWatchdogHardwareSafe2012,nagarakatteCETSCompilerEnforced2010,xiaCHERIvokeCharacterisingPointer2019}.
Drawing inspiration from software methods like 
CCured \cite{neculaCCuredTypesafeRetrofitting2002}, 
CyClone \cite{270632} 
and prior capability-based architectures 
\cite{needhamCambridgeCAPComputer1977}, 
CHERI (Capability Hardware Enhanced RISC Instructions) pioneered the approach of binding metadata directly to the ABI incompatible fat pointer 
\cite{amarCHERIoTCompleteMemory2023,chengAdaptiveCHERICompartmentalization2025,chisnallCHERIJNISinking2017,davisCheriABIEnforcingValid2019,grisenthwaiteArmMorelloEvaluation2023,watsonCHERIHybridCapabilitySystem2015,woodruffCHERIConcentratePractical2019,woodruffCHERICapabilityModel2014,xiaCHERIvokeCharacterisingPointer2019}. 
However, the CHERI model, which eschewed a centralized descriptor table, incurred a necessary overhead by using sweeping method to maintain temporal safety 
\cite{xiaCHERIvokeCharacterisingPointer2019}.

A critical turning point emerged in 2018 with the advent of microarchitectural side-channel attacks (e.g., Spectre) that leveraged speculative execution 
\cite{canellaSystematicEvaluationTransient2019,kocherSpectreAttacksExploiting2020}. 
Since software checks in the transient execution phase can still trigger microarchitectural state leakage, this threat established hardware-mandated boundary checks as the paramount research challenge in contemporary architecture.
 
Consequently, 2020 and beyond has heavily prioritized ABI-compatible pointer formats and low-overhead metadata management 
\cite{ibnziadNoFATArchitecturalSupport2021,kimHardwarebasedAlwaysOnHeap2020,xuInfatPointerHardwareassisted2021}. 
A proliferation of research, including key works such as 
MetaTableLite \cite{chenMetaTableLiteEfficientMetadata2021} and 
GPUShield \cite{leeSecuringGPURegionbased2022}, 
has effectively re-introduced the descriptor lookup approach to the security domain, marking a new phase for object-aware memory systems

\subsubsection{Memory Management}
The proliferation of big data applications and virtualization has made the Translation Lookaside Buffer (TLB) a critical performance bottleneck. 
The Range TLB \cite{basuEfficientVirtualMemory2013} 
architectural paradigm emerged as a significant response, employing a hardware/operating system co-design to boost translation efficiency by allocating contiguous physical memory to large objects and calculating the virtual-to-physical address translation via a simple VA-PA offset encoded in a range descriptor  (as shown in Figure 2-c3). This approach has generated a rich research trajectory. Early works like 
Direct Segmentation \cite{basuEfficientVirtualMemory2013} and 
RMM \cite{karakostasRedundantMemoryMappings2015} 
pioneered Range TLB architectures in L2 TLB level; 
CA-Paging \cite{alvertiEnhancingExploitingContiguity2020} 
focused on OS-level contiguous memory management using the Linux Virtual Memory Area 
(VMA) structure \cite{bovetUnderstandingLinuxKernel2008}; 
and the concept was extended to on-chip caches by 
Midgard \cite{guptaRebootingVirtualMemory2021} and to 
the GPU domain by
OASIS \cite{wangOASISObjectAwarePage2025}. 
While these studies have driven the systematic development of contiguous memory management, the reliance on finding large contiguous blocks remains a practical challenge. Consequently, some research has pivoted toward optimizing VMA-based translation without physical contiguity, such as 
DMT \cite{zhangDirectMemoryTranslation2024}
and its proposed solution of using independent linear page tables for each VMA, which is arguably the more optimal, contiguity-agnostic path for decoupling translation efficiency from the physical memory layout constraint.

\subsubsection{Memory-Centric Processing}
Recent research efforts have focused on mitigating the pervasive data movement bottleneck and improving system energy efficiency by integrating dedicated compute modules into Near Data Processing (NDP) units 
\cite{balasubramonianNearDataProcessingInsights2014,boroumandGoogleWorkloadsConsumer2018}. 
Early work implemented mechanisms akin to descriptor registers and used specialized configuration instructions to define data objects 
\cite{ahnPIMenabledInstructionsLowoverhead2015,ahnScalableProcessinginmemoryAccelerator2015}. 
A significant innovation was introduced by Baskaran et al. 
\cite{baskaranArchitectureInterfaceOffload2022}, 
who redefined the architectural interface for offloading tasks. Their design introduced explicit object identifiers and buffer identifiers as instruction operands. Crucially, they developed a Buffer Orchestrator to manage a distributed descriptor table. This architecture provided a highly efficient offload model and a clear architectural interface for low-overhead, distributed NDP accelerators, moving beyond simple data prefetching or filtering toward enabling complex, structured computation directly on the data's memory module.

\vspace{1\baselineskip}

Collectively, the body of work reviewed above, spanning from hardware-enforced memory safety mechanisms to TLB efficiency improvements and NDP accelerators, demonstrates a strong, if often implicit, convergence toward Object-Aware Memory System designs. However, a critical observation is that these innovations have largely been developed independently and in isolation within their respective domains. The next generation of computer architecture necessitates a unified, holistic object-aware memory system architecture. Future designs must synthesize the key insights from these disparate research efforts to deliver a truly integrated solution that simultaneously addresses persistent challenges in performance, security, and energy efficiency.

\section{Descriptor Addressing and Processing}
\subsection{Descriptor Addressing Dimensions}
Descriptor addressing is the mechanism that dictates the process by which memory address calculation operations obtain the relevant object's descriptor and ID information. Drawing an analogy from processor instruction addressing modes, we delineate descriptor addressing along the following dimensions:
 
\textbf{Explicit versus Implicit.} Explicit descriptor addressing modes are characterized by the instruction's source address operand or register directly containing the addressing metadata. This metadata, typically residing in dedicated descriptor registers, fat pointers, or tagged pointers, provides architectural flexibility but often necessitates non-trivial modifications to the instruction set architecture and software stack for compatibility with conventional processor designs. Conversely, implicit descriptor addressing modes maintain the standard addressing workflow, transparently mapping the current linear address to its corresponding object descriptor, thereby ensuring superior hardware and software compatibility. Implicit modes further bifurcate into lookup-table-based and sorted-table-based approaches. The lookup-table method maps each pointer address to a descriptor located at the equivalent offset within a shadow memory structure, offering high-speed descriptor retrieval at the expense of memory storage efficiency. In contrast, the sorted-table approach, utilizing  structures like B+ trees, minimizes storage overhead but introduces higher latency due to the requisite multiple memory accesses for descriptor traversal.

\textbf{Direct versus Indirect.} In the direct mode, the source address operand or register's metadata may contain the descriptor directly, or the hardware can dynamically generate the object descriptor or object ID which is also termed the descriptor generation method. Conversely, the indirect mode necessitates memory accesses to retrieve the descriptor, a category that encompasses implicit descriptor addressing and is often referred to as the identifier lookup method. Indirect modes are further categorized based on descriptor organization: distributed in-object descriptors or centralized descriptor data structure. The in-object indirect approach embeds descriptors directly within the object's data structure, thereby modifying the object's original memory layout. Centralized descriptor data structure, however, introduce coherence overhead in multithreaded environments, a challenge analogous to that encountered with page tables.

\textbf{Per-object versus per-pointer.} In the per-object model, each distinct memory object is coupled with a single, unique descriptor, meaning all memory accesses targeting that object must reference this unique descriptor. This design guarantees strict consistency across all references but often translates into substantial hardware complexity and performance overhead for descriptor management. Conversely, the per-pointer mode permits individual pointers to maintain their own local copies of the object descriptor, thereby enhancing flexibility in access control mechanisms. However, this approach introduces significant challenges concerning the timeliness of descriptor revocation and updates, and it incurs notable performance and complexity overhead when addressing temporal memory safety issues due to the need for managing distributed descriptor states.

\textbf{Fat pointers versus Tagged pointers. }The fat pointer scheme encapsulates object metadata alongside address information within an expanded pointer data structure, typically resulting in a pointer width of two or four times the native machine word size. This approach offers the capacity to carry rich metadata and deliver robust functionality but presents significant challenges in practical deployment due to its incompatibility with existing Application Binary Interfaces (ABIs) and memory models. In contrast, the tagged pointer scheme leverages unused high-order bits within the virtual address space to store concise metadata. This method maintains the original pointer width while achieving superior hardware and software compatibility, leading to its widespread adoption across both industry and academia. More recently, hybrid tagged pointer techniques have emerged to enhance flexibility by supporting a combination of multiple indirect or direct descriptor addressing modes within the pointer tag. This allows for the dynamic selection of the descriptor retrieval mechanism based on the specific characteristics and security requirements of different objects, establishing it as a key direction for future research and development in the field.

\subsection{Descriptor Addressing Modes}
In early two-dimension address space architectures prevalent from the 1950s through the 1980s, the descriptor register mechanism utilized a register-based direct descriptor addressing scheme 
\cite{coddMultiprogrammingSTRETCHFeasibility1959,mayerArchitectureBurroughsB50001982,thorntonParallelOperationControl1964}, 
where updates to the descriptor registers were managed by software (as shown in Figure 5-a1). Segmented memory systems typically employed register-indirect addressing modes (Figure 5-a2) 
\cite{childsProcessorFamilyPersonal1984,daleyVirtualMemoryProcesses1968,dennisSegmentationDesignMultiprogrammed1965}. 
To mitigate the latency associated with descriptor table lookups, these computing systems commonly incorporated Descriptor Cache in which employ the segment identifiers as the cache tag. Furthermore, 
capability-based memory protection architectures 
\cite{berstisSecurityProtectionData1980} 
and object-oriented systems 
\cite{colwellPerformanceEffectsArchitectural1988} 
generally leveraged fat-pointer-based indirect descriptor modes. These modes unified segment identifiers, intra-segment offsets, and access permissions into a single composite data type that was stored and transferred as an indivisible entity across registers and memory.

\begin{figure}[t]
	\centering
	\includegraphics[width=\linewidth]{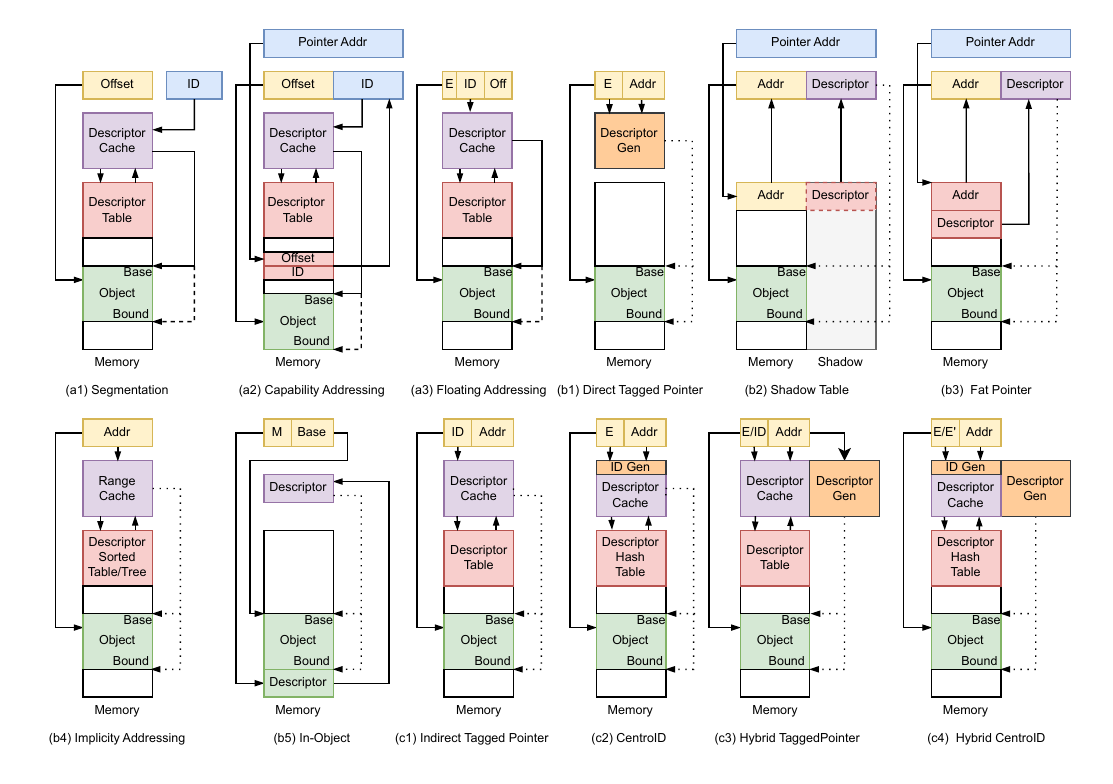}
	\caption{Descriptor Addressing Modes}
	\Description{Addressing Modes}
\end{figure}

To overcome the object capacity limitations inherent in traditional segmented memory systems, 
Dally's Catech COM \cite{dallyObjectOrientedArchitecture1985} 
introduced the innovative concept of a floating address (Figure 5-a3). This design partitions a fixed-width virtual address into three key components: a variable-length segment offset (or segment size field), a variable-length segment identifier, and an exponent field (E). The value of E dictates the bit-width of the intra-segment offset. This flexible partitioning allows fewer bits to be allocated to the offset for small data objects, consequently enabling more bits for identifiers and thereby supporting a significantly larger number of objects. This scheme elegantly accommodates segment identifiers for virtually all objects within the memory space. However, it ultimately inherited the fundamental drawbacks of segment-based addressing and failed to achieve widespread adoption, primarily due to the industry's subsequent and decisive shift towards single linear address space architectures.

Flat address space architectures constitute the prevailing design in modern Instruction Set Architectures, fundamentally decoupling addressing modes from memory protection mechanisms. Within this model, descriptors are intentionally excluded from the address calculation process. Instead, they are leveraged exclusively for bounds checking and to carry other program-invisible, cross-layer higher-level semantic metadata. Followings are an overview of the categorization of research achievements in this area.

\subsubsection{Tagged-pointer direct descriptor addressing} This addressing mode leverages pointer tags and the linear address to directly calculate object boundary information. This approach is particularly well-suited for performing spatial safety checks on a large volume of small-sized objects, although it inherently lacks comprehensive support for large or complex memory objects.As demonstrated by 
M-Machine's Guarded Pointers \cite{carterHardwareSupportFast1994} and 
Baggy Bounds Checking \cite{akritidisBaggyBoundsChecking2009} 
(illustrated in Figure 5-b1 and Figure 6-b), the prevalent Pointer Alignment method mandates that objects be allocated to minimally $2^N$-aligned memory slots. This design simplifies bounds checking to merely verifying whether the high-order bits of the address calculation result remain invariant, which substantially reduces hardware complexity. However, this method suffers from significant internal fragmentation, with a worst-case overhead approaching $2^{N-1} - 1$ bytes.In the pointer alignment method, the slot exponent $N$ is determined based on an arbitrary object size $Length$. For any post-allocation address $Pointer$ within the object, the object's boundary information $[Base, Bound]$ can be directly derived using the following expressions:
$$\text{Base} = \text{SlotBase} = \text{Pointer } \& \sim(2^N - 1)$$$$\text{Bound} = \text{Base} \ | \ (2^N - 1)$$

To mitigate the substantial internal fragmentation issue associated with prior alignment-based methods (like $\text{Baggy Bounds}$), the 
Low-Fat pointer scheme \cite{kwonLowfatPointersCompact2013} 
introduces a critical refinement: it further partitions the $2^N$-byte memory slot into $2^M$ sub-blocks of $2^E$ bytes each, where $M = N - E$ and $M$ is typically a constant. This scheme incorporates two $M$-bit fields into the pointer tag: a first block index $B$ and a last block index $T$ (assuming $T > B$).The object boundaries are then precisely computed using the following formulas:
$$\text{Base} = \text{SlotBase} \ | \ (B \ll E)$$$$\text{Bound} = \text{SlotBase} \ | \ (T \ll E) \ | \ (2^E - 1)$$

This method successfully reduces the worst-case fragmentation to  $2^{E-1} - 1$ bytes. However, it still cannot completely eliminate intra-sub-block fragmentation for very large objects. Crucially, Low-Fat necessitates integrating more metadata into the pointer, which consequently constrains the available linear address bit-width and limits the scalability of the Instruction Set Architecture. For instance, in a system with a 57-bit virtual address space, only 7 tag bits may remain, making it challenging to fully implement the Low-Fat encoding scheme. 

\begin{figure}[t]
	\centering
	\includegraphics[width=\linewidth]{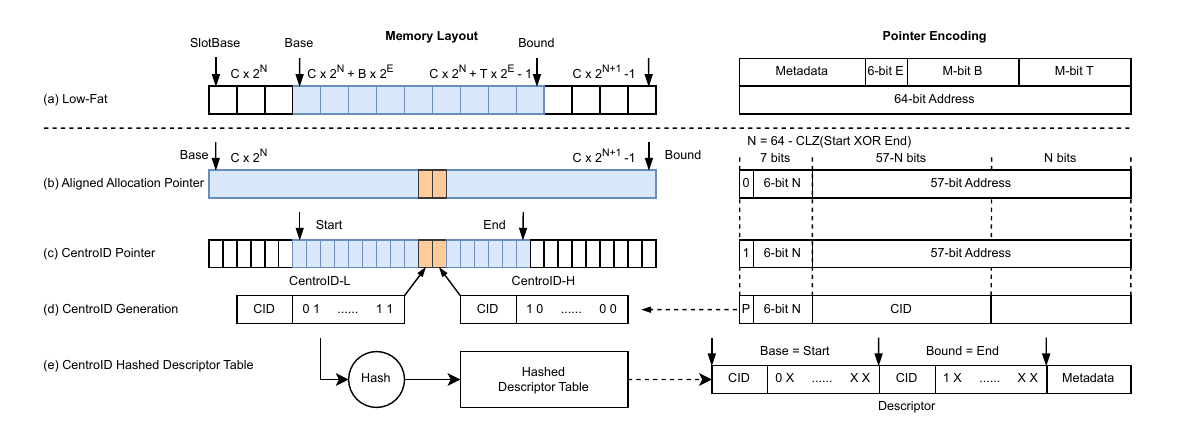}
	\caption{Tagged Pointer Encoding and the CentroID Generation}
	\Description{CentroID}
\end{figure}

\subsubsection{Fat-pointer descriptor addressing} This addressing mode has garnered significant research and practical attention, but its inherent incompatibility with existing pointer formats constrains wider adoption. As shown in Figure 5-b2, schemes like 
HardBound \cite{deviettiHardboundArchitecturalSupport2008} and 
Watchdog \cite{nagarakatteCETSCompilerEnforced2010,nagarakatteWatchdogHardwareSafe2012} 
store pointer metadata in shadow memory, facilitating rapid lookups via fixed offsets from shadow table base address. However, this method suffers from substantial memory overhead and low utility due to the sparse distribution of live pointers, rendering it primarily suitable for memory safety vulnerabilities finding rather than high-performance, fine-grained hardware memory protection. The CHERI project represents the most influential modern research in memory-safe Instruction Set Architectures. It extends the linear address space by encapsulating the memory address, object bounds, identifier, and access permissions within a capability pointer. As shown in Figure 5-b3, the early 
CHERI-256 \cite{woodruffCHERICapabilityModel2014} 
integrated a full object descriptor and permissions into a 4 address-wide fat pointer, while 
CHERI-128 \cite{grisenthwaiteArmMorelloEvaluation2023} and CHERI-64 \cite{amarCHERIoTCompleteMemory2023} 
employed the Low-Fat encoding (as shown in Figure 6-a) to compress pointers to twice the address width for 64-bit and 32-bit architectures, respectively. Nevertheless, CHERI's lack of a centralized descriptor table necessitates reliance on dynamic garbage collection to enforce temporal memory safety, which introduces a notable performance overhead. In contrast, 
Shakti-T \cite{menonShaktiTRISCVProcessor2017}
employs a fat-pointer indirect addressing model with fixed-width identifiers capable of representing all object descriptors. While this approach reduces memory overhead compared to shadow memory schemes, the per-object descriptor allocation still incurs a non-negligible memory cost.

\subsubsection{Implicit descriptor addressing} This addressing mode operates by not modifying pointer formats, instead performing descriptor lookups directly from linear addresses as shown in Figure 5-b4. Intel 
MPX \cite{oleksenkoIntelMPXExplained2018} 
implemented a radix tree-based bound table structure, analogous to page tables, yet its mechanisms for bound propagation and multi-level lookups incurred substantial performance overhead. 
BOGO \cite{zhangBOGOBuySpatial2019} 
extended MPX's bound table to additionally support temporal memory safety. 
RMM \cite{karakostasRedundantMemoryMappings2015} 
introduced the 
Range Cache \cite{tiwariSmallCacheLarge2008}
into the L2 TLB level to perform parallel range comparisons for identifying continuous physical address ranges during L1 TLB misses, while utilizing B+ trees within the operating system for efficient range queries. This hybrid approach has been widely employed in subsequent TLB and address translation research, although its hardware complexity fundamentally constrains the number of supportable objects. 
CHEx86 \cite{sharifiCHEx86ContextSensitiveEnforcement2020} 
dynamically extracts descriptor information from OS symbol tables within the microcode layer, assigning dynamic identifiers and maintaining a program-transparent descriptor table to enforce memory safety. Finally, 
NoFAT \cite{ibnziadNoFATArchitecturalSupport2021} 
employed the strategy like
HeapBounds \cite{duckHeapBoundsProtection2016} 
partitioning the address space into fixed-size regions each containing uniformly-sized objects, and introduced a hardware Memory Allocation Size Table (MAST) to retrieve object sizes based on the memory region. Overall, implicit descriptor addressing mechanisms typically introduce significant memory and performance overhead due to the necessity of complex descriptor table lookup operations.

\subsubsection{In-object descriptor addressing} This addressing mode constitutes a specialized form of indirect descriptor addressing, as illustrated in Figure 5-b5. This approach embeds object descriptor metadata within the memory object itself and encodes corresponding index information within the pointer's tag fields. Instructions utilizing this mode typically require three address components: the object base address, the intra-object descriptor offset, and the data access offset/address, making it particularly suitable for two-dimensional address formats or fat pointers. Storing metadata at a fixed offset, such as in an object header at the start, eliminates the need for explicit descriptor offset metadata. 
HotPad \cite{tsaiRethinkingMemoryHierarchy2018} 
used pointer alignment to compute object bounds and stored metadata indices in the object's first word, supporting object-granular cache optimization and garbage collection for modern programming languages, with subsequent work extending its use to object-granular compression. The 
ALEXIA \cite{krishnakumarALEXIAProcessorLightweight2019} 
processor employed fat pointers containing object identifiers and base addresses while storing bounds and matching identifiers within the objects; upon deallocation, the internal identifier is cleared, enabling temporal safety verification via dynamic pointer-object identifier matching. 
ViK \cite{choViKPracticalMitigation2022} 
similarly used aligned tagged pointers with another object ID tag, and store the same ID at fixed object location for temporal safety, and leveraged static analysis to identify pointers requiring hardware protection owing to limited tag capacity. 
C3 \cite{lemayCryptographicCapabilityComputing2021} 
introduced a stateless memory-safe tagged pointer scheme that encodes bounds into cryptographic keys. It allocates each object to a minimally-fitting 2N sized slot, using the slot's center point combined with pointer tags as unique object identifiers. By storing the exponent N in high-order pointer bits and generating encryption keys from invariant address bitfields, C3 enables data encryption and integrity verification during decryption by recovering encryption the key to validate object integrity.

\subsubsection{Tagged-pointer indirect descriptor addressing} This addressing mode enhances memory safety and address translation by integrating metadata indices within pointer tag fields and maintaining descriptor tables in memory, as depicted in Figure 5-c1. The 
AOS scheme \cite{kimHardwarebasedAlwaysOnHeap2020} 
leverages ARM Pointer Authentication Codes 
(PAC) \cite{liljestrandPACItPointer2019} 
to generate unique keys for querying an extensible hash-based bounds table, incorporating a dedicated bounds cache to mitigate lookup latency—though PAC bit limitations necessitate hash collision resolution. 
HeapCheck \cite{saileshwarHeapCheckLowcostHardware2022} 
similarly encodes bounds table indices directly into pointer tags, facing comparable constraints on the number of supported objects. Inspired by the RMM,
FlexPointer \cite{chenFlexPointerFastAddress2023} 
innovatively applies tagged-pointer indirect addressing to address translation by utilizing a compact 13-bit interval index table for the largest objects. It stores these interval indices in pointer tags and employs set-associative structures, termed RangeTLB, to increase the Range Cache entry counts in RMM. As shown in Figure 5-c2, FlexPointer further explores the feasibility of ID-based hash table.

\subsubsection{Tagged-pointer hybrid descriptor addressing} This addressing mode integrates a mode selector field within the pointer tag to dynamically choose the appropriate descriptor addressing mechanism based on object types (Figure 5-c3). 
In-Fat pointer\cite{xuInfatPointerHardwareassisted2021} 
adapts the 
EffectiveSan \cite{duckEffectiveSanTypeMemory2018} 
approach, utilizing in-object addressing to achieve fine-grained sub-object protection. It incorporates a policy selector in the pointer tag that supports three indirect metadata addressing modes: local offset resolution, configuration register-based sub-heap management, and global table lookup. Meanwhile, 
MetaTableLite \cite{chenMetaTableLiteEfficientMetadata2021} and 
GPUShield \cite{leeSecuringGPURegionbased2022} 
addresses memory protection by combining tagged-pointer direct addressing and indirect addressing to a small descriptor table. While this offers structural simplicity and operational flexibility, its fixed-size descriptor table in indirect mode inherently constrains scalability. Future object-aware memory systems should leverage this foundational mode to further reduce the tag bit-width and eliminate the limitation of descriptor table entry counts. Section 5 proposes a hybrid CentroID solution that directly addresses these critical challenges (Figure 5-c4).

\subsection{Descriptor Generation Unit}
To ensure comprehensive memory safety and system reliability while enabling precise cross-layer semantic information transfer, object-aware architectures must incorporate explicit pointer types, extended pointer arithmetic instructions, and descriptor processing instructions. This requires guaranteeing the legality and integrity of pointers throughout their construction and propagation. The principles established by the CHERI project—least privilege, monotonicity, and explicit operation—provide an ideal blueprint for future Instruction Set Architecture extensions in this domain. Specifically, we recommend a hybrid tagged-pointer scheme coupled with indirect addressing based on a centralized descriptor table. Furthermore, hardware-enforced mandatory security checks must be applied to all address calculation micro-operations and descriptor handling from system startup, ensuring full trust in pointer propagation.

\begin{figure}[t]
	\centering
	\includegraphics[width=\linewidth]{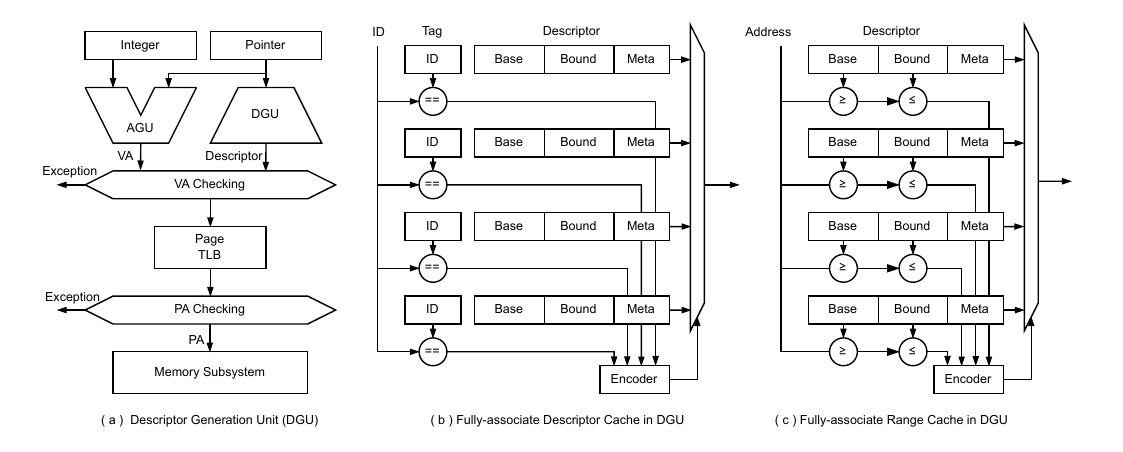}
	\caption{a) Descriptor Generation Unit and b) Descriptor Cache c) Range Cache}
	\Description{AGU}
\end{figure}

As illustrated in Figure 7a, in the Address Generation Unit (AGU) of an object-aware system, object Descriptor Generation Unit (DGU) proceeds in parallel with address computation. The AGU calculates the target effective address using the pointer and offset while DGU simultaneously retrieving the object descriptor based on the active descriptor addressing mode. A dedicated virtual address check module then checks whether the target address falls within the bounds and permissions defined by the descriptor. If the check passes, the target address and descriptor are forwarded to the memory subsystem; otherwise, an exception is raised.

For indirect descriptor addressing modes, the descriptor processing unit must fetch descriptors from memory, necessitating a dedicated descriptor cache to reduce average access latency. Two mainstream caching mechanisms exist: First, the identifier-based conventional structure, termed Descriptor Cache which uses object identifiers as tags and descriptors as data (Figure 7b), backed by memory-resident lookup or hash tables. Second, the 
Range Cache mechanism \cite{fengBarreChordEfficient2024,guptaRebootingVirtualMemory2021,karakostasRedundantMemoryMappings2015,kumarMETALCachingMultilevel2024,tiwariSmallCacheLarge2008}, 
which eliminates the need for tag storage by employing fully associative parallel comparison for queries (Figure 7c). However, Range Cache typically supports a limited number of entries (e.g., 16) and still incurs lookup overhead when relying on sorted tables or search trees in memory for misses. Identifier-based caches and descriptor tables offer superior performance and hardware efficiency in general, while Range Cache remains suitable only for specialized scenarios lacking explicit object identifiers.

\section{CENTROID APPROACH}
\subsection{Minimal Tag Bit-Width Scheme}
Designing a tagged-pointer scheme utilizing minimal pointer tags (e.g., 7 bits) presents both significant application potential and considerable architectural challenges. In conventional descriptor mechanisms for memory objects, a descriptor typically encompasses at least dual-address boundary information, and attaching full descriptor structures to every object introduces substantial metadata storage overhead. Furthermore, the variable size distribution of memory objects complicates the design of both efficient descriptor organization and high-performance lookup algorithms. Based on studies of dynamic object behavior, descriptor architecture design should adhere to three key principles: first, support the elastic management of massive concurrent objects to accommodate diverse application requirements; second, employ lightweight metadata structures for the predominant small, short-lived objects to optimize system overhead; and third, utilize enhanced descriptors with extended metadata and more sophisticated lookup mechanisms for the minority of large, long-lived objects. Such a holistic scheme must balance storage efficiency, access performance, and security control to establish a viable path for object-aware computing architectures.

Inspired by the prior work on hybrid tagged-pointer encoding 
\cite{chenMetaTableLiteEfficientMetadata2021}, 
Aligned Allocation Pointer as utilized in M-Machine \cite{carterHardwareSupportFast1994}, 
Baggy Bounds Checking \cite{akritidisBaggyBoundsChecking2009}, and 
LMI \cite{leeLetMeInStillEmploying2025}, 
can be effectively applied to the majority of small objects (under 1KB) that primarily require bounds checks. This approach incurs acceptable memory overhead according to 
mimalloc \cite{leijenMimallocFreeList2019} 
evaluation and requires only 6 tag bits in a 64-bit address space. While Low-Fat encoding (Figure 7a) reduce internal fragmentation, they rely on fat pointers with ABI incompatibility issues 
\cite{amarCHERIoTCompleteMemory2023,grisenthwaiteArmMorelloEvaluation2023,woodruffCHERIConcentratePractical2019}, 
whereas the Aligned Allocation Pointer offers satisfactory fragmentation behavior for small objects (Figure 7b). 

For special memory objects, including very large objects, and those requiring precise bounds or extended cross-layer metadata, the indirect descriptor addressing is appropriate. As in 
FlexPointer \cite{chenFlexPointerFastAddress2023}, 
a tagged pointer encoding using only 6 bits pointer tag to generate unique identifiers (Figure 7c) is presented, referred to here as the CentroID Pointer.

\subsection{CentroID Pointer Encoding}

The CentroID mechanism provides an efficient methodology for object identification and boundary description in modern object-aware architectures. Based on the principle of minimal aligned allocation slot, this scheme can generate the unique identifiers for any memory object. This unique encoding method discussed in prior research 
\cite{chenFlexPointerFastAddress2023,lemayCryptographicCapabilityComputing2021,namFRAMERTaggedpointerCapability2019,unterguggenbergerCryptographicallyEnforcedMemory2023}, 
is detailed below. 

For any memory object spanning the virtual address range [Start, End], where Start < End, the aligned allocation principle ensures the existence of a minimal 2N bytes aligned memory slot, termed N-Slot, that fully contains the object. Given a machine word size W, a floating-point bitfield CID, where its bit-width is W – N – log2(W) -1, exists such that all virtual addresses within the object's range share identical values. By analyzing address boundary characteristics, the key parameters are efficiently computed as: this calculation leverages the binary property that the Nth least significant bit of Start must be zero and the Nth least significant bit of End must be one, enabling rapid determination of the alignment granularity N via XOR operation of Start and End  and the count leading zero function 
\cite{namFRAMERTaggedpointerCapability2019}. 

It is evident that each $2^N$-aligned memory slot contains two special midpoints, referred to as CentroID-L and CentroID-H. Specifically, $\text{CentroID-L}$ is defined by setting its $\text{N}$-th least significant bit ($\text{LSB}$) to zero and its lower $(\text{N}-1)$ bits to all-ones. Conversely, $\text{CentroID-H}$ is defined by setting its $\text{N}$-th $\text{LSB}$ to one and its lower $(\text{N}-1)$ bits to all-zeros, as illustrated in Figure 7d . From a geometric perspective, these two points are located at the "centroid" positions of the memory slot, effectively bisecting it.A significant advantage of selecting these two points as identifiers is that the memory slot's size exponent $N$ can be easily and efficiently determined from $\text{CentroID-L}$ or $\text{CentroID-H}$ using the Count Trailing Ones ($\text{CTO}$) or Count Trailing Zeros ($\text{CTZ}$) operation, respectively.Assuming the slot base is calculated as $\text{SlotBase} = \text{Pointer} \ \& \ \sim(2^N - 1)$, the two $\text{CentroID}$ midpoints are derived using the following bitwise expressions:$$\text{CentroID-L} = \text{SlotBase} \ | \ (2^{N-1} - 1)$$$$\text{CentroID-H} = \text{SlotBase} \ | \ \sim(2^{N-1} - 1)$$

Mathematically, it can be proven that within a specific N-Slot, only one valid memory object can encompass both CentroID-L and CentroID-H, irrespective of that object's Start and End addresses. Conversely, any given memory object can arbitrarily select one object ID as its unique identifier. Generally, based on the resulting allocated address range, a small object may be assigned a relatively large exponent N. This situation should be avoided wherever possible, so that the exponent N naturally corresponds to the actual size of the memory object. 

The CentroID-based object-aware system employs a hybrid pointer encoding format (Figure 7b, 7c). The highest pointer bit serves as a mode selector bit: zero indicates an Aligned Allocation Pointer, while one denotes a CentroID Pointer. In 64-bit architectures, the subsequent 6-bit field stores the alignment memory slot exponent N. 

The descriptor generation unit performs operations based on the pointer mode: for Aligned Allocation Pointer, it directly calculates object boundary information; for CentroID pointers, it retrieves object metadata through a descriptor cache tagged with CentroID. On a descriptor cache miss, as illustrated in Figure 7e, the system queries a descriptor hash table in memory where the CentroID as the hash key. This layered processing mechanism maintains hardware efficiency while providing a unified management framework for memory objects of different scales, establishing a theoretical foundation for building scalable object-aware memory systems. 

The CentroID approach maintains pointer width compatibility, substantially easing the adaptation of existing operating systems and applications without requiring changes to core algorithms or data structures—only the generation of a tagged pointer based on origin allocation results. Implementation is tailored to the allocation context: for system initialization and kmalloc (which typically uses power-of-two sizing), the Alignment Allocation Pointer mode is applied directly; for mmap allocations managed via VMAs, the CentroID Pointer mode is employed, augmented by a CentroID descriptor hash table atop the VMA structure. In the case of standard malloc heap allocations, the CentroID is computed externally from the malloc metadata and the pointer tag updated accordingly, while for user-managed memory, developers are advised to employ the same mechanism used for heap allocations via the malloc interface to ensure consistent CentroID computation and application.

\subsection{Multi-CentroID Descriptor Scheme}

\begin{figure}[t]
	\centering
	\includegraphics[width=\linewidth]{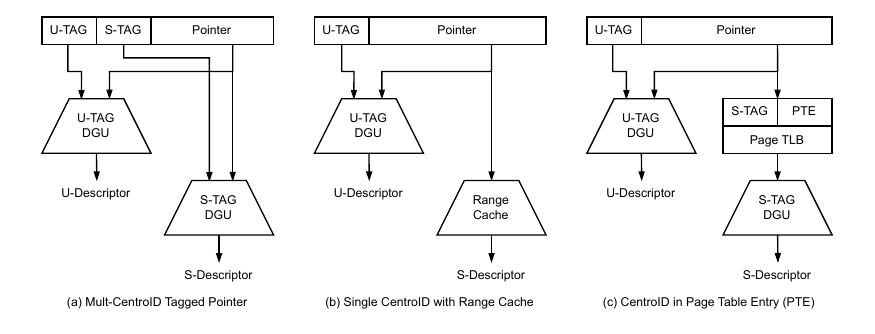}
	\caption{Multi-CentroID Descriptor Processing}
	\Description{Unified}
\end{figure}

To simultaneously support memory protection and memory management using a single tagged-pointer encoding scheme, an object-aware memory system must incorporate a multi-level descriptor mechanism. 

In a typical memory allocation process, nested object relationships often exist, where a child object is allocated internally from a parent object. As a design principle, reverse lookup of the parent object's metadata based solely on the child object's information is generally prohibited. If the focus is strictly on memory protection, ensuring the child object's memory safety is often sufficient to guarantee the overall security of the parent object, in which case a single-tag pointer mechanism meets the requirement.

However, in the practical allocation workflow, the system first allocates a page-granularity parent object via system calls ($\text{e.g.}$, $\text{mmap}$), requiring a system-level descriptor to support memory management functions like address translation. Subsequently, small-sized child objects are derived from the parent using library functions ($\text{e.g.}$, $\text{malloc}$), and these objects require a user-level descriptor for fine-grained memory safety protection. In this scenario, the conventional single-tagged pointer mechanism fails to support the memory system's ability to dynamically lookup the descriptor of the parent object from its child object metadata.

Three alternative solutions can be naturally proposed to address this problem, as illustrated in Figure 8, where U-TAG denotes the tag in user mode and S-TAG denotes the tag in system mode.

\textbf{Multi-Tag Pointer Approach}. The first approach (Figure 8a) employs a dual-tag encoding scheme where two distinct pointer tag fields concurrently incorporate the metadata for the parent and child objects, respectively, processed by two independent $\text{DGUs}$. This design is highly suitable for scenarios where an ample pointer tag bitfield is available, such as a 48-bit virtual address in 64-bit architecture providing a 16-bit tag bitfield capable of encoding two distinct tags. However, this approach inherently further diminishes the available virtual address width, potentially constraining the usable address space such as 57 bits virtual address. A more critical disadvantage is the potential for information leakage concerning system-level state, as the combined encoding of the S-TAG and U-TAG could expose system metadata to user mode, introducing a vulnerability exploitable by an attacker.

\textbf{Range Cache Approach}. The second approach (Figure 8b) employs a Range Cache parallel comparison mechanism to manage the limited set of system-level objects. This method utilizes fully-associative lookups to quickly match a child object's address with its corresponding parent object's descriptor. While this solution necessitates higher hardware complexity due to the fully-associative lookup structure, its primary advantage is that it entirely avoids consuming valuable pointer tag bitfields, preserving the full pointer width for other purposes.

\textbf{Page Table Entry Tagging Approach}. The third approach (Figure 8c) leverages the existing address translation infrastructure by integrating the S-TAG field directly into the Page Table Entries ($\text{PTEs}$). This strategy inherently associates system-level object descriptors with the address translation mechanism, which is supplemented by $\text{DGU}$. Under typical execution, these extended $\text{PTEs}$ directly supply the system-level object $\text{CentroID}$s. For performance-critical large objects, a dedicated Descriptor Cache employed to enable low-latency lookup. This hybrid design achieves an optimal trade-off among storage efficiency, performance, and hardware complexity, providing a robust foundation for constructing practical object-aware memory systems. Future research should focus on optimizing the encoding density of $\text{CentroID}$s within page tables and exploring more efficient hierarchical descriptor lookup mechanisms.

The three schemes each possess distinct trade-offs regarding applicability and implementation cost. The Multi-Tag Pointer scheme is best suited for scenarios that support nested objects within the same privilege level, provided ample tag space is available. Conversely, the Range Cache mechanism is a compelling option for environments lacking page-based virtual memory, such as traditional embedded systems, where it utilizes dedicated hardware without modifying the tagged-pointer encoding. The PTE Tagging Scheme, despite offering the most desirable trade-off among efficiency, complexity, and storage, requires extensive modifications to both the operating system and TLB hardware, posing the highest barrier to adoption.

\subsection{CentroID-based Object-Aware Memory Systems}
The CentroID-based hybrid tagged pointer encoding and multi-CentroID descriptor generation method provide a comprehensive descriptor handling mechanism for implementing general-purpose object-aware memory systems. This approach can generate address bounds and unique identifiers for all memory objects using minimal pointer tag bitfield. The address bounds enable enforcing mandatory address validation to ensure complete spatial memory safety. Crucially, it enables even Aligned Allocation Pointers to deterministically derive their associated CentroID. These universal identifiers can further enhance temporal memory safety, memory management and memory processing optimizations.

The CentroID-based unified scheme presents a promising avenue for enforcing temporal memory safety, particularly in mitigating Use-After-Free (UAF) vulnerabilities, with and comprehensive lifecycle management. By integrating the CentroID mechanism into existing protection frameworks, such as 
PTAuth \cite{263812} or 
ViK \cite{choViKPracticalMitigation2022}, 
the inherent limitations on the number of tracked objects can be overcome. Alternatively, introducing CentroID into safe memory allocation in systems like 
MarkUs \cite{ainsworthMarkUsDropinUseafterfree2020} or 
FFMalloc \cite{263880} 
can establish a more efficient object lookup mechanism, consequently improving overall system performance.

Furthermore, the CentroID-based descriptor cache and hash table mechanisms offer a compelling, high-performance alternative to conventional descriptor storage structures. Specifically, they can effectively supersede established components, such as hardware Range Caches or software sorted trees, often utilized in contemporary memory management research efforts like 
Flexpoint \cite{chenFlexPointerFastAddress2023}, 
Midgard \cite{guptaRebootingVirtualMemory2021}, 
DMT \cite{zhangDirectMemoryTranslation2024}, 
and Jord \cite{liSingleAddressSpaceFaaSJord2025}. 
This substitution leads to substantial enhancements in descriptor lookup performance and a tangible reduction in associated overhead.

Moreover, the CentroID tagged pointer mechanism presents several compelling implementation pathways for significantly advancing memory-centric processing. First, this mechanism facilitates the direct and seamless substitution of existing object identifier schemes within memory-centric instruction set \cite{baskaranArchitectureInterfaceOffload2022}. 
This capability simultaneously enhances both programmability and flexibility for developers, all while ensuring full backward compatibility with current architectural models. Second, fundamental, high-frequency library routines, such as memory movement, which are conventionally implemented using lengthy instruction sequences, can be transformed into dedicated, CentroID-based instructions. Within this paradigm, the hardware automatically intercepts the CentroID, generates the corresponding object descriptors, enabling more secure and efficient memory processing.

\section{Related Research Directions}
This section surveys several existing research directions that inherently incorporate object-aware characteristics. We present only a few representative works in each area to illustrate the architectural necessity for such mechanisms, many of which leverage Object ID or descriptor-like mechanisms. The eventual development of a general object-aware architectural framework promises to substantially elevate the efficacy and scope of these specialized research avenues.

\subsection{Cache Hierarchies}
Many high-performance cache replacement and prefetching strategies leverage Program Counter (PC) correlation mechanisms \cite{ayersClassifyingMemoryAccess2020,bakhshalipourBingoSpatialData2019,beraPythiaCustomizableHardware2021,bhatiaPerceptronbasedPrefetchFiltering2019,hashemiLearningMemoryAccess2018,khanSamplingDeadBlock2010,navarro-torresBertiAccurateLocalDelta2022,pakalapatiBouquetInstructionPointers2020,peledSemanticLocalityContextbased2015,shiApplyingDeepLearning2019,shiHierarchicalNeuralModel2021,wuSHiPSignaturebasedHit2011}. 
Since a specific PC in the instruction stream strictly correlates with accesses to a singular memory object, these PC-correlated cache management strategies are inherently object-aware. However, incorporating PCs into physically-tagged cache hierarchies introduce significant hardware design complexity and overhead. Furthermore, prefetching without explicit hardware object bounds checking creates potential side-channel attack risks \cite{chenPREFETCHXCrossCoreCacheAgnostic2024,grussPrefetchSideChannelAttacks2016}. 
To address these limitations, future object-aware memory systems should propagate object descriptors (at minimum, custom-defined page size) and an Object ID to the cache hierarchy. This foundational information enables object-based cache behavior learning, along with in-bounds prefetching and compression optimizations. 
On the other hand, several research efforts have proposed cross-layer methodologies, spanning system interfaces, instruction set architectures (ISAs), and hardware innovations, to explicitly communicate high-level semantic metadata into the cache hierarchy. For instance, 
XMem \cite{vijaykumarLocalityDescriptorHolistic2018,vijaykumarCaseRicherCrossLayer2018,vijaykumarMetaSysPracticalOpensource2022} 
introduced a holistic software-hardware co-design approach to assign identifiers (referred to as Atoms) to specific memory objects, binding these identifiers to fixed-size memory regions via an Atom Address Map mechanism akin to Range Cache. Furthermore, 
HotPad \cite{tsaiRethinkingMemoryHierarchy2018,tsaiCompressObjectsNot2019} 
utilized tagged-pointer and in-object descriptor techniques to design and implement an object-granularity on-chip hierarchy, enabling advanced cache system management and garbage collection optimizations support for modern programming languages.

\subsection{Unified Virtual Memory}
Unified Virtual Memory (UVM) establishes a shared address space between CPUs and GPUs, enabling transparent data movement and sharing \cite{gangulyInterplayHardwarePrefetcher2019,jungDeepUMTensorMigration2023,linDrGPUMGuidingMemory2023,zhangG10EnablingEfficient2023}. 
This abstraction significantly reduces programming complexity and improves overall system utilization. The concept of UVM is now evolving toward a truly unified system model that encompasses all forms of heterogeneous processing units, including 
accelerators \cite{haoSupportingAddressTranslation2017}, 
intelligent I/O controllers \cite{amitIOMMUStrategiesMitigating2010}, 
and processing-in-memory (PIM) architectures \cite{leePIMMMUMemoryManagement2024}, 
within a single virtual memory framework. The ultimate goal is to create a computing paradigm in which data from any memory technology can be accessed uniformly across the entire hierarchy, irrespective of its physical location, whether local or remote, on-chip or off-chip \cite{liPondCXLBasedMemory2023,limDisaggregatedMemoryExpansion2009,marufTPPTransparentPage2023,meswaniHeterogeneousMemoryArchitectures2015}. 
A critical aspect involves migrating pages with strong data locality to nearby, high-speed memory regions through software-managed policies. Consequently, the core technical challenge in UVM design lies in developing robust mechanisms to track, and quantitatively characterize page-level locality. 

This persistent challenge in efficient data placement and migration has driven the adoption of higher-level abstractions. specifically range descriptors and object identifiers. For instance, 
Capuchin \cite{pengCapuchinTensorbasedGPU2020} 
leverages tensor IDs for runtime monitoring and prediction of tensor access patterns, thereby optimizing page migration decisions. Architecturally, AMD’s Shared Virtual Memory (SVM) diverges from conventional page table management by utilizing a continuous page-range model, a design that inherently facilitates object-level and descriptor-based memory management. Following this direction, related research such as Copper et al. on prefetching and replacement strategies optimized for 
the SVM model \cite{cooperSharedVirtualMemory2024}. 
Lin et al.'s proposal \cite{linForestAccessawareGPU2025} 
for a range-based prefetching scheme tailored for NVIDIA GPU’s Unified Memory (UM) ecosystem.

\subsection{Persistent Memory}
Persistent Memory (PM) systems introduce a high-density, non-volatile layer between DRAM and SSDs in the storage hierarchy. A crucial issue for PM is the correct association between residual data objects following a system crash and the new memory allocation objects upon system restart. To solve this, researchers 
\cite{chenEfficientSupportPosition2017,wangHardwareSupportedPersistent2017}  
have proposed deconstructing virtual addresses into object identifiers and offsets. These identifiers, along with segment-table-like structures, are used to reconstruct the original object addresses. Taking an alternative route, the Twizzler operating system \cite{bittmanTwizzlerDatacentricOS2021a}
maximizes the bit width available for offset addressing by encoding compact table indexes in the pointer tag. This compact table containing the full 128-bit Object IDs then provide an indirection to the complete object table. The CentroID method can be further enhanced by incorporating a 
Floating Address approach \cite{dallyObjectOrientedArchitecture1985} 
efficiently supporting the maintenance of PM consistency.

\subsection{Architectural Supports for Modern Programming Languages}
Modern high-level programming languages, such as Java, Go, and Python, significantly enhance developer productivity and code reliability through built-in features like memory safety guarantees and object-oriented data abstraction. However, the implementation of these high-level semantics on current commodity hardware and operating systems incurs substantial runtime performance overhead. This inefficiency is fundamentally rooted in a structural impedance mismatch between the language's execution model and the underlying memory system architecture. Specifically, two mechanisms become critical bottlenecks \cite{caoYinYangPower2012,kimTypedArchitecturesArchitectural2017,shullNoMapSpeedingUpJavaScript2019,tsaiRethinkingMemoryHierarchy2018}: 
the mandatory dynamic checking, which consumes excessive CPU cycles by introducing numerous branch operations that disrupt instruction pipelines; and the overhead associated with garbage collection (GC), where runtime interventions severely constrain the application's latency and throughput. Object-Aware Architectures represent a pivotal innovation that narrows the semantic gap between modern programming languages and the underlying hardware. By directly embedding high-level language constructs—such as object structures, pointers/references, and type metadata—into the microarchitecture, these designs fundamentally enhance the security and performance of modern software execution \cite{davisCheriABIEnforcingValid2019,kimTypedArchitecturesArchitectural2017,liSingleAddressSpaceFaaSJord2025,tsaiRethinkingMemoryHierarchy2018,wangMementoArchitecturalSupport2023}.

\subsection{Next-Generation 128-bit Computer Architecture}
Scaling big data applications presents a formidable challenge, as dataset sizes are projected to swiftly approach the 64-bit virtual address ceiling. This limit, equating to exabyte ($ \text{EiB} ,  2^{60} $) scale, will soon be surpassed as data move rapidly into the zettabyte ($ \text{ZiB} ,  2^{70} $) and yottabyte ($ \text{YiB} ,  2^{80} $) scales. Consequently, the academic community and industry are actively exploring 
Zettascale supercomputing \cite{gomesExascaleParadigmShift2023,liaoMovingExascaleZettascale2018,su11InnovationNext2023} 
and investigating 128-bit computer architectures \cite{deshpandePractical128BitGeneral2023}. 
The CentroID-based hybrid tagged-pointer scheme can offer a promising solution for these future systems. The 128-bit virtual address provides sufficient unused bits to encode multiple CentroID tags, which in turn enables a more sophisticated, multi-tiered descriptor-based object-aware memory system.

\section{Conclusion}
This survey has systematically charted the architectural evolution of descriptor-based object-aware memory systems, establishing a unified analytical framework through a novel taxonomy of descriptor addressing modes. We have demonstrated that the fundamental inefficiencies and vulnerabilities pervasive in contemporary memory architectures stem from a persistent semantic gap—the inherent disconnect between software's object-oriented view and the underlying hardware's block-oriented operation. By generating unique object identifiers and precise bounds from minimal pointer tags, CentroID provides a practical, high-performance path toward hardware-enforced spatial and temporal memory safety without sacrificing the performance of large-object management or the integrity of existing ABI. Ultimately, CentroID hybrid tagged-pointer encoding and multi-tiered descriptor scheme offer a blueprint for the next generation of computing architecture, laying the groundwork for a truly integrated and unified object-aware memory system that addresses the persistent challenges of security, performance, and efficiency simultaneously.

\begin{acks}
The completion of this work was significantly inspired by the dedication and hard work of the Ph.D. and Master's students at Peking University who participated in research projects on the memory safety, virtual memory and memory hierarchy. Especially thanks to Dongwei Chen, Daliang Xu, Kang Sun, Honglei Zhang, Zhijun Yu, Mi Liu, Yidi Wang, Haoqian Chen for their valuable contributions. 
\end{acks}

\bibliographystyle{ACM-Reference-Format}
\bibliography{CentroID}

@inproceedings{263812,
  title = {{{PTAuth}}: {{Temporal}} Memory Safety via Robust Points-to Authentication},
  booktitle = {30th {{USENIX}} Security Symposium ({{USENIX}} Security 21)},
  author = {{farkhani}, Reza Mirzazade and Ahmadi, Mansour and Lu, Long},
  year = 2021,
  month = aug,
  pages = {1037--1054},
  publisher = {USENIX Association},
  isbn = {978-1-939133-24-3}
}

@inproceedings{263880,
  title = {Preventing {{Use-After-Free}} Attacks with Fast Forward Allocation},
  booktitle = {30th {{USENIX}} Security Symposium ({{USENIX}} Security 21)},
  author = {Wickman, Brian and Hu, Hong and Yun, Insu and Jang, DaeHee and Lim, JungWon and Kashyap, Sanidhya and Kim, Taesoo},
  year = 2021,
  month = aug,
  pages = {2453--2470},
  publisher = {USENIX Association},
  isbn = {978-1-939133-24-3}
}

@inproceedings{270632,
  title = {Cyclone: A Safe Dialect of {{C}}},
  booktitle = {2002 {{USENIX}} Annual Technical Conference ({{USENIX ATC}} 02)},
  author = {Jim, Trevor and Morrisett, Greg and Grossman, Dan and Hicks, Michael and Cheney, James and Wang, Yanling},
  year = 2002,
  month = jun,
  publisher = {USENIX Association},
  address = {Monterey, CA}
}

@inproceedings{ahnPIMenabledInstructionsLowoverhead2015,
  title = {{{PIM-enabled}} Instructions: A Low-Overhead, Locality-Aware Processing-in-Memory Architecture},
  shorttitle = {{{PIM-enabled}} Instructions},
  booktitle = {Proceedings of the 42nd {{Annual International Symposium}} on {{Computer Architecture}}},
  author = {Ahn, Junwhan and Yoo, Sungjoo and Mutlu, Onur and Choi, Kiyoung},
  year = 2015,
  month = jun,
  pages = {336--348},
  publisher = {ACM},
  address = {Portland Oregon},
  doi = {10.1145/2749469.2750385},
  urldate = {2025-08-09},
  isbn = {978-1-4503-3402-0},
  langid = {english}
}

@inproceedings{ahnScalableProcessinginmemoryAccelerator2015,
  title = {A Scalable Processing-in-Memory Accelerator for Parallel Graph Processing},
  booktitle = {Proceedings of the 42nd {{Annual International Symposium}} on {{Computer Architecture}}},
  author = {Ahn, Junwhan and Hong, Sungpack and Yoo, Sungjoo and Mutlu, Onur and Choi, Kiyoung},
  year = 2015,
  month = jun,
  pages = {105--117},
  publisher = {ACM},
  address = {Portland Oregon},
  doi = {10.1145/2749469.2750386},
  urldate = {2025-08-24},
  isbn = {978-1-4503-3402-0},
  langid = {english}
}

@inproceedings{ainsworthMarkUsDropinUseafterfree2020,
  title = {{{MarkUs}}: {{Drop-in}} Use-after-Free Prevention for Low-Level Languages},
  shorttitle = {{{MarkUs}}},
  booktitle = {2020 {{IEEE Symposium}} on {{Security}} and {{Privacy}} ({{SP}})},
  author = {Ainsworth, Sam and Jones, Timothy M.},
  year = 2020,
  month = may,
  pages = {578--591},
  publisher = {IEEE},
  address = {San Francisco, CA, USA},
  doi = {10.1109/SP40000.2020.00058},
  urldate = {2025-10-17},
  copyright = {https://ieeexplore.ieee.org/Xplorehelp/downloads/license-information/IEEE.html},
  isbn = {978-1-7281-3497-0}
}

@inproceedings{akritidisBaggyBoundsChecking2009,
  title = {Baggy Bounds Checking: An Efficient and Backwards-Compatible Defense against out-of-Bounds Errors},
  booktitle = {Proceedings of the 18th {{Conference}} on {{USENIX Security Symposium}}},
  author = {Akritidis, Periklis and Costa, Manuel and Castro, Miguel and Hand, Steven},
  year = 2009,
  month = aug,
  series = {{{SSYM}}'09},
  pages = {51--66},
  publisher = {USENIX Association},
  address = {USA}
}

@inproceedings{alvertiEnhancingExploitingContiguity2020,
  title = {Enhancing and {{Exploiting Contiguity}} for {{Fast Memory Virtualization}}},
  booktitle = {2020 {{ACM}}/{{IEEE}} 47th {{Annual International Symposium}} on {{Computer Architecture}} ({{ISCA}})},
  author = {Alverti, Chloe and Psomadakis, Stratos and Karakostas, Vasileios and Gandhi, Jayneel and Nikas, Konstantinos and Goumas, Georgios and Koziris, Nectarios},
  year = 2020,
  month = may,
  pages = {515--528},
  publisher = {IEEE},
  address = {Valencia, Spain},
  doi = {10.1109/ISCA45697.2020.00050},
  urldate = {2025-08-11},
  copyright = {https://doi.org/10.15223/policy-029},
  isbn = {978-1-7281-4661-4}
}

@inproceedings{amarCHERIoTCompleteMemory2023,
  title = {{{CHERIoT}}: {{Complete Memory Safety}} for {{Embedded Devices}}},
  shorttitle = {{{CHERIoT}}},
  booktitle = {56th {{Annual IEEE}}/{{ACM International Symposium}} on {{Microarchitecture}}},
  author = {Amar, Saar and Chisnall, David and Chen, Tony and Filardo, Nathaniel Wesley and Laurie, Ben and Liu, Kunyan and Norton, Robert and Moore, Simon W. and Tao, Yucong and Watson, Robert N. M. and Xia, Hongyan},
  year = 2023,
  month = oct,
  pages = {641--653},
  publisher = {ACM},
  address = {Toronto ON Canada},
  doi = {10.1145/3613424.3614266},
  urldate = {2025-08-12},
  isbn = {979-8-4007-0329-4},
  langid = {english}
}

@inproceedings{amitIOMMUStrategiesMitigating2010,
  title = {{{IOMMU}}: Strategies for Mitigating the {{IOTLB}} Bottleneck},
  booktitle = {Proceedings of the 2010 {{International Conference}} on {{Computer Architecture}}},
  author = {Amit, Nadav and {Ben-Yehuda}, Muli and Yassour, Ben-Ami},
  year = 2010,
  month = jun,
  series = {{{ISCA}}'10},
  pages = {256--274},
  publisher = {Springer-Verlag},
  address = {Berlin, Heidelberg},
  doi = {10.1007/978-3-642-24322-6_22},
  isbn = {978-3-642-24321-9}
}

@inproceedings{ayersClassifyingMemoryAccess2020,
  title = {Classifying {{Memory Access Patterns}} for {{Prefetching}}},
  booktitle = {Proceedings of the {{Twenty-Fifth International Conference}} on {{Architectural Support}} for {{Programming Languages}} and {{Operating Systems}}},
  author = {Ayers, Grant and Litz, Heiner and Kozyrakis, Christos and Ranganathan, Parthasarathy},
  year = 2020,
  month = mar,
  pages = {513--526},
  publisher = {ACM},
  address = {Lausanne Switzerland},
  doi = {10.1145/3373376.3378498},
  urldate = {2025-08-11},
  isbn = {978-1-4503-7102-5},
  langid = {english}
}

@inproceedings{bakhshalipourBingoSpatialData2019,
  title = {Bingo {{Spatial Data Prefetcher}}},
  booktitle = {2019 {{IEEE International Symposium}} on {{High Performance Computer Architecture}} ({{HPCA}})},
  author = {Bakhshalipour, Mohammad and Shakerinava, Mehran and {Lotfi-Kamran}, Pejman and {Sarbazi-Azad}, Hamid},
  year = 2019,
  month = feb,
  pages = {399--411},
  publisher = {IEEE},
  address = {Washington, DC, USA},
  doi = {10.1109/HPCA.2019.00053},
  urldate = {2025-08-12},
  copyright = {https://ieeexplore.ieee.org/Xplorehelp/downloads/license-information/IEEE.html},
  isbn = {978-1-7281-1444-6}
}

@article{balasubramonianNearDataProcessingInsights2014,
  title = {Near-{{Data Processing}}: {{Insights}} from a {{MICRO-46 Workshop}}},
  shorttitle = {Near-{{Data Processing}}},
  author = {Balasubramonian, Rajeev and Chang, Jichuan and Manning, Troy and Moreno, Jaime H. and Murphy, Richard and Nair, Ravi and Swanson, Steven},
  year = 2014,
  month = jul,
  journal = {IEEE Micro},
  volume = {34},
  number = {4},
  pages = {36--42},
  issn = {0272-1732, 1937-4143},
  doi = {10.1109/MM.2014.55},
  urldate = {2025-08-09},
  copyright = {https://ieeexplore.ieee.org/Xplorehelp/downloads/license-information/IEEE.html}
}

@inproceedings{baskaranArchitectureInterfaceOffload2022,
  title = {An Architecture Interface and Offload Model for Low-Overhead, near-Data, Distributed Accelerators},
  booktitle = {2022 55th {{IEEE}}/{{ACM International Symposium}} on {{Microarchitecture}} ({{MICRO}})},
  author = {Baskaran, Saambhavi and Kandemir, Mahmut Taylan and Sampson, Jack},
  year = 2022,
  month = oct,
  pages = {1160--1177},
  publisher = {IEEE},
  address = {Chicago, IL, USA},
  doi = {10.1109/MICRO56248.2022.00083},
  urldate = {2025-08-09},
  copyright = {https://doi.org/10.15223/policy-029},
  isbn = {978-1-6654-6272-3}
}

@inproceedings{basuEfficientVirtualMemory2013,
  title = {Efficient Virtual Memory for Big Memory Servers},
  booktitle = {Proceedings of the 40th {{Annual International Symposium}} on {{Computer Architecture}}},
  author = {Basu, Arkaprava and Gandhi, Jayneel and Chang, Jichuan and Hill, Mark D. and Swift, Michael M.},
  year = 2013,
  month = jun,
  pages = {237--248},
  publisher = {ACM},
  address = {Tel-Aviv Israel},
  doi = {10.1145/2485922.2485943},
  urldate = {2025-08-09},
  isbn = {978-1-4503-2079-5},
  langid = {english}
}

@inproceedings{beraPythiaCustomizableHardware2021,
  title = {Pythia: {{A Customizable Hardware Prefetching Framework Using Online Reinforcement Learning}}},
  shorttitle = {Pythia},
  booktitle = {{{MICRO-54}}: 54th {{Annual IEEE}}/{{ACM International Symposium}} on {{Microarchitecture}}},
  author = {Bera, Rahul and Kanellopoulos, Konstantinos and Nori, Anant and Shahroodi, Taha and Subramoney, Sreenivas and Mutlu, Onur},
  year = 2021,
  month = oct,
  pages = {1121--1137},
  publisher = {ACM},
  address = {Virtual Event Greece},
  doi = {10.1145/3466752.3480114},
  urldate = {2025-08-09},
  isbn = {978-1-4503-8557-2},
  langid = {english}
}

@inproceedings{berstisSecurityProtectionData1980,
  title = {Security and Protection of Data in the {{IBM System}}/38},
  booktitle = {Proceedings of the 7th Annual Symposium on {{Computer Architecture}}  - {{ISCA}} '80},
  author = {Berstis, Viktors},
  year = 1980,
  month = may,
  pages = {245--252},
  publisher = {ACM Press},
  address = {La Baule, United States},
  doi = {10.1145/800053.801932},
  urldate = {2025-08-10},
  copyright = {https://www.acm.org/publications/policies/copyright\_policy\#Background},
  langid = {english}
}

@inproceedings{bhatiaPerceptronbasedPrefetchFiltering2019,
  title = {Perceptron-Based Prefetch Filtering},
  booktitle = {Proceedings of the 46th {{International Symposium}} on {{Computer Architecture}}},
  author = {Bhatia, Eshan and Chacon, Gino and Pugsley, Seth and Teran, Elvira and Gratz, Paul V. and Jim{\'e}nez, Daniel A.},
  year = 2019,
  month = jun,
  pages = {1--13},
  publisher = {ACM},
  address = {Phoenix Arizona},
  doi = {10.1145/3307650.3322207},
  urldate = {2025-08-12},
  isbn = {978-1-4503-6669-4},
  langid = {english}
}

@article{bittmanTwizzlerDatacentricOS2021a,
  title = {Twizzler: {{A}} {{{\emph{Data-centric}}}} {{OS}} for {{Non-volatile Memory}}},
  shorttitle = {Twizzler},
  author = {Bittman, Daniel and Alvaro, Peter and Mehra, Pankaj and Long, Darrell D. E. and Miller, Ethan L.},
  year = 2021,
  month = may,
  journal = {ACM Transactions on Storage},
  volume = {17},
  number = {2},
  pages = {1--31},
  issn = {1553-3077, 1553-3093},
  doi = {10.1145/3454129},
  urldate = {2025-10-20},
  langid = {english}
}

@inproceedings{bodikABCDEliminatingArray2000,
  title = {{{ABCD}}: Eliminating Array Bounds Checks on Demand},
  shorttitle = {{{ABCD}}},
  booktitle = {Proceedings of the {{ACM SIGPLAN}} 2000 Conference on {{Programming}} Language Design and Implementation},
  author = {Bod{\'i}k, Rastislav and Gupta, Rajiv and Sarkar, Vivek},
  year = 2000,
  month = may,
  pages = {321--333},
  publisher = {ACM},
  address = {Vancouver British Columbia Canada},
  doi = {10.1145/349299.349342},
  urldate = {2025-08-26},
  isbn = {978-1-58113-199-4},
  langid = {english}
}

@inproceedings{boroumandGoogleWorkloadsConsumer2018,
  title = {Google {{Workloads}} for {{Consumer Devices}}: {{Mitigating Data Movement Bottlenecks}}},
  shorttitle = {Google {{Workloads}} for {{Consumer Devices}}},
  booktitle = {Proceedings of the {{Twenty-Third International Conference}} on {{Architectural Support}} for {{Programming Languages}} and {{Operating Systems}}},
  author = {Boroumand, Amirali and Ghose, Saugata and Kim, Youngsok and Ausavarungnirun, Rachata and Shiu, Eric and Thakur, Rahul and Kim, Daehyun and Kuusela, Aki and Knies, Allan and Ranganathan, Parthasarathy and Mutlu, Onur},
  year = 2018,
  month = mar,
  pages = {316--331},
  publisher = {ACM},
  address = {Williamsburg VA USA},
  doi = {10.1145/3173162.3173177},
  urldate = {2025-08-09},
  isbn = {978-1-4503-4911-6},
  langid = {english}
}

@book{bovetUnderstandingLinuxKernel2008,
  title = {Understanding the {{Linux Kernel}}},
  author = {Bovet, Daniel P. and Cesati, Marco},
  year = 2008,
  edition = {3rd ed},
  publisher = {O'Reilly Media, Inc},
  address = {Sebastopol},
  isbn = {978-0-596-00565-8},
  langid = {english}
}

@inproceedings{canellaSystematicEvaluationTransient2019,
  title = {A {{Systematic Evaluation}} of {{Transient Execution Attacks}} and {{Defenses}}},
  booktitle = {28th {{USENIX Security Symposium}} ({{USENIX Security}} 19)},
  author = {Canella, Claudio and Bulck, Jo Van and Schwarz, Michael and Lipp, Moritz and von Berg, Benjamin and Ortner, Philipp and Piessens, Frank and Evtyushkin, Dmitry and Gruss, Daniel},
  year = 2019,
  month = aug,
  pages = {249--266},
  publisher = {USENIX Association},
  address = {Santa Clara, CA},
  isbn = {978-1-939133-06-9}
}

@inproceedings{caoYinYangPower2012,
  title = {The {{Yin}} and {{Yang}} of Power and Performance for Asymmetric Hardware and Managed Software},
  booktitle = {2012 39th {{Annual International Symposium}} on {{Computer Architecture}} ({{ISCA}})},
  author = {Cao, Ting and Blackburn, Stephen M and Gao, Tiejun and McKinley, Kathryn S},
  year = 2012,
  month = jun,
  pages = {225--236},
  publisher = {IEEE},
  address = {Portland, OR, USA},
  doi = {10.1109/ISCA.2012.6237020},
  urldate = {2025-08-26},
  isbn = {978-1-4673-0476-4 978-1-4673-0475-7 978-1-4673-0473-3 978-1-4673-0474-0}
}

@inproceedings{carterHardwareSupportFast1994,
  title = {Hardware Support for Fast Capability-Based Addressing},
  booktitle = {Proceedings of the Sixth International Conference on {{Architectural}} Support for Programming Languages and Operating Systems},
  author = {Carter, Nicholas P. and Keckler, Stephen W. and Dally, William J.},
  year = 1994,
  month = nov,
  pages = {319--327},
  publisher = {ACM},
  address = {San Jose California USA},
  doi = {10.1145/195473.195579},
  urldate = {2025-08-10},
  isbn = {978-0-89791-660-8},
  langid = {english}
}

@inproceedings{chenEfficientSupportPosition2017,
  title = {Efficient Support of Position Independence on Non-Volatile Memory},
  booktitle = {Proceedings of the 50th {{Annual IEEE}}/{{ACM International Symposium}} on {{Microarchitecture}}},
  author = {Chen, Guoyang and Zhang, Lei and Budhiraja, Richa and Shen, Xipeng and Wu, Youfeng},
  year = 2017,
  month = oct,
  pages = {191--203},
  publisher = {ACM},
  address = {Cambridge Massachusetts},
  doi = {10.1145/3123939.3124543},
  urldate = {2025-08-12},
  isbn = {978-1-4503-4952-9},
  langid = {english}
}

@article{chenFlexPointerFastAddress2023,
  title = {{{FlexPointer}}: {{Fast Address Translation Based}} on {{Range TLB}} and {{Tagged Pointers}}},
  shorttitle = {{{FlexPointer}}},
  author = {Chen, Dongwei and Tong, Dong and Yang, Chun and Yi, Jiangfang and Cheng, Xu},
  year = 2023,
  month = jun,
  journal = {ACM Transactions on Architecture and Code Optimization},
  volume = {20},
  number = {2},
  pages = {1--24},
  issn = {1544-3566, 1544-3973},
  doi = {10.1145/3579854},
  urldate = {2025-08-09},
  langid = {english}
}

@inproceedings{chengAdaptiveCHERICompartmentalization2025,
  title = {Adaptive {{CHERI Compartmentalization}} for {{Heterogeneous Accelerators}}},
  booktitle = {Proceedings of the 52nd {{Annual International Symposium}} on {{Computer Architecture}}},
  author = {Cheng, Jianyi and Markettos, A. Theodore and Joannou, Alexandre and Metzger, Paul and Naylor, Matthew and Rugg, Peter and Jones, Timothy M.},
  year = 2025,
  month = jun,
  pages = {2002--2016},
  publisher = {ACM},
  address = {Tokyo Japan},
  doi = {10.1145/3695053.3731062},
  urldate = {2025-09-23},
  isbn = {979-8-4007-1261-6},
  langid = {english}
}

@inproceedings{chenMetaTableLiteEfficientMetadata2021,
  title = {{{MetaTableLite}}: {{An Efficient Metadata Management Scheme}} for {{Tagged-Pointer-Based Spatial Safety}}},
  shorttitle = {{{MetaTableLite}}},
  booktitle = {2021 {{IEEE}} 39th {{International Conference}} on {{Computer Design}} ({{ICCD}})},
  author = {Chen, Dongwei and Tong, Dong and Yang, Chun and Cheng, Xu},
  year = 2021,
  month = oct,
  pages = {208--211},
  publisher = {IEEE},
  address = {Storrs, CT, USA},
  doi = {10.1109/ICCD53106.2021.00042},
  urldate = {2025-08-11},
  copyright = {https://ieeexplore.ieee.org/Xplorehelp/downloads/license-information/IEEE.html},
  isbn = {978-1-6654-3219-1}
}

@inproceedings{chenPREFETCHXCrossCoreCacheAgnostic2024,
  title = {{{PREFETCHX}}: {{Cross-Core Cache-Agnostic Prefetcher-based Side-Channel Attacks}}},
  shorttitle = {{{PREFETCHX}}},
  booktitle = {2024 {{IEEE International Symposium}} on {{High-Performance Computer Architecture}} ({{HPCA}})},
  author = {Chen, Yun and Hajiabadi, Ali and Pei, Lingfeng and Carlson, Trevor E.},
  year = 2024,
  month = mar,
  pages = {395--408},
  publisher = {IEEE},
  address = {Edinburgh, United Kingdom},
  doi = {10.1109/HPCA57654.2024.00037},
  urldate = {2025-08-26},
  copyright = {https://doi.org/10.15223/policy-029},
  isbn = {979-8-3503-9313-2}
}

@article{childsProcessorFamilyPersonal1984,
  title = {A Processor Family for Personal Computers},
  author = {Childs, R.E. and Crawford, J. and House, D.L. and Noyce, R.N.},
  year = 1984,
  month = mar,
  journal = {Proceedings of the IEEE},
  volume = {72},
  number = {3},
  pages = {363--376},
  issn = {0018-9219},
  doi = {10.1109/PROC.1984.12867},
  urldate = {2025-08-11},
  copyright = {https://ieeexplore.ieee.org/Xplorehelp/downloads/license-information/IEEE.html}
}

@inproceedings{chisnallCHERIJNISinking2017,
  title = {{{CHERI JNI}}: {{Sinking}} the {{Java Security Model}} into the {{C}}},
  shorttitle = {{{CHERI JNI}}},
  booktitle = {Proceedings of the {{Twenty-Second International Conference}} on {{Architectural Support}} for {{Programming Languages}} and {{Operating Systems}}},
  author = {Chisnall, David and Davis, Brooks and Gudka, Khilan and Brazdil, David and Joannou, Alexandre and Woodruff, Jonathan and Markettos, A. Theodore and Maste, J. Edward and Norton, Robert and Son, Stacey and Roe, Michael and Moore, Simon W. and Neumann, Peter G. and Laurie, Ben and Watson, Robert N.M.},
  year = 2017,
  month = apr,
  pages = {569--583},
  publisher = {ACM},
  address = {Xi'an China},
  doi = {10.1145/3037697.3037725},
  urldate = {2025-10-10},
  isbn = {978-1-4503-4465-4},
  langid = {english}
}

@inproceedings{choViKPracticalMitigation2022,
  title = {{{ViK}}: Practical Mitigation of Temporal Memory Safety Violations through Object {{ID}} Inspection},
  shorttitle = {{{ViK}}},
  booktitle = {Proceedings of the 27th {{ACM International Conference}} on {{Architectural Support}} for {{Programming Languages}} and {{Operating Systems}}},
  author = {Cho, Haehyun and Park, Jinbum and Oest, Adam and Bao, Tiffany and Wang, Ruoyu and Shoshitaishvili, Yan and Doup{\'e}, Adam and Ahn, Gail-Joon},
  year = 2022,
  month = feb,
  pages = {271--284},
  publisher = {ACM},
  address = {Lausanne Switzerland},
  doi = {10.1145/3503222.3507780},
  urldate = {2025-08-12},
  isbn = {978-1-4503-9205-1},
  langid = {english}
}

@inproceedings{coburnNVHeapsMakingPersistent2011,
  title = {{{NV-Heaps}}: Making Persistent Objects Fast and Safe with next-Generation, Non-Volatile Memories},
  shorttitle = {{{NV-Heaps}}},
  booktitle = {Proceedings of the Sixteenth International Conference on {{Architectural}} Support for Programming Languages and Operating Systems},
  author = {Coburn, Joel and Caulfield, Adrian M. and Akel, Ameen and Grupp, Laura M. and Gupta, Rajesh K. and Jhala, Ranjit and Swanson, Steven},
  year = 2011,
  month = mar,
  pages = {105--118},
  publisher = {ACM},
  address = {Newport Beach California USA},
  doi = {10.1145/1950365.1950380},
  urldate = {2025-10-27},
  isbn = {978-1-4503-0266-1},
  langid = {english}
}

@article{coddMultiprogrammingSTRETCHFeasibility1959,
  title = {Multiprogramming {{STRETCH}}: Feasibility Considerations},
  shorttitle = {Multiprogramming {{STRETCH}}},
  author = {Codd, E. F. and Lowry, E. S. and McDonough, E. and Scalzi, C. A.},
  year = 1959,
  month = nov,
  journal = {Communications of the ACM},
  volume = {2},
  number = {11},
  pages = {13--17},
  issn = {0001-0782, 1557-7317},
  doi = {10.1145/368481.368502},
  urldate = {2025-09-22},
  langid = {english}
}

@article{colwellPerformanceEffectsArchitectural1988,
  title = {Performance Effects of Architectural Complexity in the {{Intel}} 432},
  author = {Colwell, Robert P. and Gehringer, Edward F. and Jensen, E. Douglas},
  year = 1988,
  month = aug,
  journal = {ACM Transactions on Computer Systems},
  volume = {6},
  number = {3},
  pages = {296--339},
  issn = {0734-2071, 1557-7333},
  doi = {10.1145/45059.214411},
  urldate = {2025-08-10},
  langid = {english}
}

@inproceedings{cooperSharedVirtualMemory2024,
  title = {Shared {{Virtual Memory}}: {{Its Design}} and {{Performance Implications}} for {{Diverse Applications}}},
  shorttitle = {Shared {{Virtual Memory}}},
  booktitle = {Proceedings of the 38th {{ACM International Conference}} on {{Supercomputing}}},
  author = {Cooper, Bennett and Scogland, Thomas Rw and Ge, Rong},
  year = 2024,
  month = may,
  pages = {26--37},
  publisher = {ACM},
  address = {Kyoto Japan},
  doi = {10.1145/3650200.3656608},
  urldate = {2025-08-11},
  isbn = {979-8-4007-0610-3},
  langid = {english}
}

@article{daleyVirtualMemoryProcesses1968,
  title = {Virtual Memory, Processes, and Sharing in {{MULTICS}}},
  author = {Daley, Robert C. and Dennis, Jack B.},
  year = 1968,
  month = may,
  journal = {Communications of the ACM},
  volume = {11},
  number = {5},
  pages = {306--312},
  issn = {0001-0782, 1557-7317},
  doi = {10.1145/363095.363139},
  urldate = {2025-08-11},
  langid = {english}
}

@article{dallyObjectOrientedArchitecture1985,
  title = {An Object Oriented Architecture},
  author = {Dally, William J. and Kajiya, James T.},
  year = 1985,
  month = jun,
  journal = {ACM SIGARCH Computer Architecture News},
  volume = {13},
  number = {3},
  pages = {154--161},
  issn = {0163-5964},
  doi = {10.1145/327070.327151},
  urldate = {2025-09-22},
  copyright = {https://www.acm.org/publications/policies/copyright\_policy\#Background},
  langid = {english}
}

@article{dassharmaIntroductionComputeExpress2024,
  title = {An {{Introduction}} to the {{Compute Express Link}} ({{CXL}}) {{Interconnect}}},
  author = {Das Sharma, Debendra and Blankenship, Robert and Berger, Daniel},
  year = 2024,
  month = nov,
  journal = {ACM Computing Surveys},
  volume = {56},
  number = {11},
  pages = {1--37},
  issn = {0360-0300, 1557-7341},
  doi = {10.1145/3669900},
  urldate = {2025-10-21},
  langid = {english}
}

@inproceedings{davisCheriABIEnforcingValid2019,
  title = {{{CheriABI}}: {{Enforcing Valid Pointer Provenance}} and {{Minimizing Pointer Privilege}} in the {{POSIX C Run-time Environment}}},
  shorttitle = {{{CheriABI}}},
  booktitle = {Proceedings of the {{Twenty-Fourth International Conference}} on {{Architectural Support}} for {{Programming Languages}} and {{Operating Systems}}},
  author = {Davis, Brooks and Watson, Robert N. M. and Richardson, Alexander and Neumann, Peter G. and Moore, Simon W. and Baldwin, John and Chisnall, David and Clarke, Jessica and Filardo, Nathaniel Wesley and Gudka, Khilan and Joannou, Alexandre and Laurie, Ben and Markettos, A. Theodore and Maste, J. Edward and Mazzinghi, Alfredo and Napierala, Edward Tomasz and Norton, Robert M. and Roe, Michael and Sewell, Peter and Son, Stacey and Woodruff, Jonathan},
  year = 2019,
  month = apr,
  pages = {379--393},
  publisher = {ACM},
  address = {Providence RI USA},
  doi = {10.1145/3297858.3304042},
  urldate = {2025-09-30},
  isbn = {978-1-4503-6240-5},
  langid = {english}
}

@article{denningVirtualMemory1970,
  title = {Virtual {{Memory}}},
  author = {Denning, Peter J.},
  year = 1970,
  month = sep,
  journal = {ACM Computing Surveys},
  volume = {2},
  number = {3},
  pages = {153--189},
  issn = {0360-0300, 1557-7341},
  doi = {10.1145/356571.356573},
  urldate = {2025-08-10},
  langid = {english}
}

@article{dennisProgrammingSemanticsMultiprogrammed1966,
  title = {Programming Semantics for Multiprogrammed Computations},
  author = {Dennis, Jack B. and Van Horn, Earl C.},
  year = 1966,
  month = mar,
  journal = {Communications of the ACM},
  volume = {9},
  number = {3},
  pages = {143--155},
  issn = {0001-0782, 1557-7317},
  doi = {10.1145/365230.365252},
  urldate = {2025-08-10},
  langid = {english}
}

@article{dennisSegmentationDesignMultiprogrammed1965,
  title = {Segmentation and the {{Design}} of {{Multiprogrammed Computer Systems}}},
  author = {Dennis, Jack B.},
  year = 1965,
  month = oct,
  journal = {Journal of the ACM},
  volume = {12},
  number = {4},
  pages = {589--602},
  issn = {0004-5411, 1557-735X},
  doi = {10.1145/321296.321310},
  urldate = {2025-08-10},
  langid = {english}
}

@article{deshpandePractical128BitGeneral2023,
  title = {Toward {{Practical}} 128-{{Bit General Purpose Microarchitectures}}},
  author = {Deshpande, Chandana S. and Perais, Arthur and P{\'e}trot, Fr{\'e}d{\'e}ric},
  year = 2023,
  month = jul,
  journal = {IEEE Computer Architecture Letters},
  volume = {22},
  number = {2},
  pages = {81--84},
  issn = {1556-6056, 1556-6064, 2473-2575},
  doi = {10.1109/LCA.2023.3287762},
  urldate = {2025-10-08},
  copyright = {https://ieeexplore.ieee.org/Xplorehelp/downloads/license-information/IEEE.html}
}

@inproceedings{deviettiHardboundArchitecturalSupport2008,
  title = {Hardbound: Architectural Support for Spatial Safety of the {{C}} Programming Language},
  shorttitle = {Hardbound},
  booktitle = {Proceedings of the 13th International Conference on {{Architectural}} Support for Programming Languages and Operating Systems},
  author = {Devietti, Joe and Blundell, Colin and Martin, Milo M. K. and Zdancewic, Steve},
  year = 2008,
  month = mar,
  pages = {103--114},
  publisher = {ACM},
  address = {Seattle WA USA},
  doi = {10.1145/1346281.1346295},
  urldate = {2025-08-27},
  isbn = {978-1-59593-958-6},
  langid = {english}
}

@inproceedings{dhawanArchitecturalSupportSoftwareDefined2015,
  title = {Architectural {{Support}} for {{Software-Defined Metadata Processing}}},
  booktitle = {Proceedings of the {{Twentieth International Conference}} on {{Architectural Support}} for {{Programming Languages}} and {{Operating Systems}}},
  author = {Dhawan, Udit and Hritcu, Catalin and Rubin, Raphael and Vasilakis, Nikos and Chiricescu, Silviu and Smith, Jonathan M. and Knight, Thomas F. and Pierce, Benjamin C. and DeHon, Andre},
  year = 2015,
  month = mar,
  pages = {487--502},
  publisher = {ACM},
  address = {Istanbul Turkey},
  doi = {10.1145/2694344.2694383},
  urldate = {2025-08-11},
  isbn = {978-1-4503-2835-7},
  langid = {english}
}

@inproceedings{duckEffectiveSanTypeMemory2018,
  title = {{{EffectiveSan}}: Type and Memory Error Detection Using Dynamically Typed {{C}}/{{C}}++},
  shorttitle = {{{EffectiveSan}}},
  booktitle = {Proceedings of the 39th {{ACM SIGPLAN Conference}} on {{Programming Language Design}} and {{Implementation}}},
  author = {Duck, Gregory J. and Yap, Roland H. C.},
  year = 2018,
  month = jun,
  pages = {181--195},
  publisher = {ACM},
  address = {Philadelphia PA USA},
  doi = {10.1145/3192366.3192388},
  urldate = {2025-09-02},
  isbn = {978-1-4503-5698-5},
  langid = {english}
}

@inproceedings{duckHeapBoundsProtection2016,
  title = {Heap Bounds Protection with Low Fat Pointers},
  booktitle = {Proceedings of the 25th {{International Conference}} on {{Compiler Construction}}},
  author = {Duck, Gregory J. and Yap, Roland H. C.},
  year = 2016,
  month = mar,
  pages = {132--142},
  publisher = {ACM},
  address = {Barcelona Spain},
  doi = {10.1145/2892208.2892212},
  urldate = {2025-09-02},
  isbn = {978-1-4503-4241-4},
  langid = {english}
}

@inproceedings{dulloorDataTieringHeterogeneous2016,
  title = {Data Tiering in Heterogeneous Memory Systems},
  booktitle = {Proceedings of the {{Eleventh European Conference}} on {{Computer Systems}}},
  author = {Dulloor, Subramanya R. and Roy, Amitabha and Zhao, Zheguang and Sundaram, Narayanan and Satish, Nadathur and Sankaran, Rajesh and Jackson, Jeff and Schwan, Karsten},
  year = 2016,
  month = apr,
  pages = {1--16},
  publisher = {ACM},
  address = {London United Kingdom},
  doi = {10.1145/2901318.2901344},
  urldate = {2025-10-11},
  isbn = {978-1-4503-4240-7},
  langid = {english}
}

@article{fabryCapabilitybasedAddressing1974,
  title = {Capability-Based Addressing},
  author = {Fabry, R. S.},
  year = 1974,
  month = jul,
  journal = {Communications of the ACM},
  volume = {17},
  number = {7},
  pages = {403--412},
  issn = {0001-0782, 1557-7317},
  doi = {10.1145/361011.361070},
  urldate = {2025-08-10},
  langid = {english}
}

@inproceedings{fengBarreChordEfficient2024,
  title = {Barre {{Chord}}: {{Efficient Virtual Memory Translation}} for {{Multi-Chip-Module GPUs}}},
  shorttitle = {Barre {{Chord}}},
  booktitle = {2024 {{ACM}}/{{IEEE}} 51st {{Annual International Symposium}} on {{Computer Architecture}} ({{ISCA}})},
  author = {Feng, Yuan and Na, Seonjin and Kim, Hyesoon and Jeon, Hyeran},
  year = 2024,
  month = jun,
  pages = {834--847},
  publisher = {IEEE},
  address = {Buenos Aires, Argentina},
  doi = {10.1109/ISCA59077.2024.00065},
  urldate = {2025-09-29},
  copyright = {https://doi.org/10.15223/policy-029},
  isbn = {979-8-3503-2658-1}
}

@article{feustelAdvantagesTaggedArchitecture1973,
  title = {On {{The Advantages}} of {{Tagged Architecture}}},
  author = {Feustel, Edward A.},
  year = 1973,
  month = jul,
  journal = {IEEE Transactions on Computers},
  volume = {C-22},
  number = {7},
  pages = {644--656},
  issn = {0018-9340},
  doi = {10.1109/TC.1973.5009130},
  urldate = {2025-08-10},
  copyright = {https://ieeexplore.ieee.org/Xplorehelp/downloads/license-information/IEEE.html}
}

@inproceedings{gangulyInterplayHardwarePrefetcher2019,
  title = {Interplay between Hardware Prefetcher and Page Eviction Policy in {{CPU-GPU}} Unified Virtual Memory},
  booktitle = {Proceedings of the 46th {{International Symposium}} on {{Computer Architecture}}},
  author = {Ganguly, Debashis and Zhang, Ziyu and Yang, Jun and Melhem, Rami},
  year = 2019,
  month = jun,
  pages = {224--235},
  publisher = {ACM},
  address = {Phoenix Arizona},
  doi = {10.1145/3307650.3322224},
  urldate = {2025-10-09},
  isbn = {978-1-4503-6669-4},
  langid = {english}
}

@article{gholamiAIMemoryWall2024,
  title = {{{AI}} and {{Memory Wall}}},
  author = {Gholami, Amir and Yao, Zhewei and Kim, Sehoon and Hooper, Coleman and Mahoney, Michael W. and Keutzer, Kurt},
  year = 2024,
  month = may,
  journal = {IEEE Micro},
  volume = {44},
  number = {3},
  pages = {33--39},
  issn = {0272-1732, 1937-4143},
  doi = {10.1109/MM.2024.3373763},
  urldate = {2025-08-09},
  copyright = {https://ieeexplore.ieee.org/Xplorehelp/downloads/license-information/IEEE.html}
}

@inproceedings{glaserSystemDesignComputer1965,
  title = {System Design of a Computer for Time Sharing Applications},
  booktitle = {Proceedings of the {{November}} 30--{{December}} 1, 1965, Fall Joint Computer Conference, Part {{I}} on {{XX}} - {{AFIPS}} '65 ({{Fall}}, Part {{I}})},
  author = {Glaser, E. L. and Couleur, J. F. and Oliver, G. A.},
  year = 1965,
  month = nov,
  pages = {197},
  publisher = {ACM Press},
  address = {Las Vegas, Nevada},
  doi = {10.1145/1463891.1463913},
  urldate = {2025-09-22},
  copyright = {https://www.acm.org/publications/policies/copyright\_policy\#Background},
  langid = {english}
}

@inproceedings{gomesExascaleParadigmShift2023,
  title = {Beyond {{Exascale}}: {{A Paradigm}} Shift for {{AI}} and {{HPC}}},
  shorttitle = {Beyond {{Exascale}}},
  booktitle = {2023 {{International Electron Devices Meeting}} ({{IEDM}})},
  author = {Gomes, W.},
  year = 2023,
  month = dec,
  pages = {1--4},
  issn = {2156-017X},
  doi = {10.1109/IEDM45741.2023.10413754},
  urldate = {2025-09-23}
}

@inproceedings{gonzalezProfilingHyperscaleBig2023,
  title = {Profiling {{Hyperscale Big Data Processing}}},
  booktitle = {Proceedings of the 50th {{Annual International Symposium}} on {{Computer Architecture}}},
  author = {Gonzalez, Abraham and Kolli, Aasheesh and Khan, Samira and Liu, Sihang and Dadu, Vidushi and Karandikar, Sagar and Chang, Jichuan and Asanovic, Krste and Ranganathan, Parthasarathy},
  year = 2023,
  month = jun,
  pages = {1--16},
  publisher = {ACM},
  address = {Orlando FL USA},
  doi = {10.1145/3579371.3589082},
  urldate = {2025-08-09},
  isbn = {979-8-4007-0095-8},
  langid = {english}
}

@article{grahamProtectionInformationProcessing1968,
  title = {Protection in an Information Processing Utility},
  author = {Graham, Robert M.},
  year = 1968,
  month = may,
  journal = {Communications of the ACM},
  volume = {11},
  number = {5},
  pages = {365--369},
  issn = {0001-0782, 1557-7317},
  doi = {10.1145/363095.363146},
  urldate = {2025-09-22},
  langid = {english}
}

@inproceedings{grahamProtectionPrinciplesPractice1971,
  title = {Protection: Principles and Practice},
  shorttitle = {Protection},
  booktitle = {Proceedings of the {{November}} 16-18, 1971, Fall Joint Computer Conference on - {{AFIPS}} '71 ({{Fall}})},
  author = {Graham, G. Scott and Denning, Peter J.},
  year = 1971,
  month = nov,
  pages = {417},
  publisher = {ACM Press},
  address = {Las Vegas, Nevada},
  doi = {10.1145/1478873.1478928},
  urldate = {2025-08-11},
  copyright = {https://www.acm.org/publications/policies/copyright\_policy\#Background},
  langid = {english}
}

@article{grisenthwaiteArmMorelloEvaluation2023,
  title = {The {{Arm Morello Evaluation Platform}}---{{Validating CHERI-Based Security}} in a {{High-Performance System}}},
  author = {Grisenthwaite, Richard and Barnes, Graeme and Watson, Robert N. M. and Moore, Simon W. and Sewell, Peter and Woodruff, Jonathan},
  year = 2023,
  month = may,
  journal = {IEEE Micro},
  volume = {43},
  number = {3},
  pages = {50--57},
  issn = {0272-1732, 1937-4143},
  doi = {10.1109/MM.2023.3264676},
  urldate = {2025-09-01},
  copyright = {https://ieeexplore.ieee.org/Xplorehelp/downloads/license-information/IEEE.html}
}

@inproceedings{grussPrefetchSideChannelAttacks2016,
  title = {Prefetch {{Side-Channel Attacks}}: {{Bypassing SMAP}} and {{Kernel ASLR}}},
  shorttitle = {Prefetch {{Side-Channel Attacks}}},
  booktitle = {Proceedings of the 2016 {{ACM SIGSAC Conference}} on {{Computer}} and {{Communications Security}}},
  author = {Gruss, Daniel and Maurice, Cl{\'e}mentine and Fogh, Anders and Lipp, Moritz and Mangard, Stefan},
  year = 2016,
  month = oct,
  pages = {368--379},
  publisher = {ACM},
  address = {Vienna Austria},
  doi = {10.1145/2976749.2978356},
  urldate = {2025-08-26},
  isbn = {978-1-4503-4139-4},
  langid = {english}
}

@inproceedings{guptaRebootingVirtualMemory2021,
  title = {Rebooting {{Virtual Memory}} with {{Midgard}}},
  booktitle = {2021 {{ACM}}/{{IEEE}} 48th {{Annual International Symposium}} on {{Computer Architecture}} ({{ISCA}})},
  author = {Gupta, Siddharth and Bhattacharyya, Atri and Oh, Yunho and Bhattacharjee, Abhishek and Falsafi, Babak and Payer, Mathias},
  year = 2021,
  month = jun,
  pages = {512--525},
  publisher = {IEEE},
  address = {Valencia, Spain},
  doi = {10.1109/ISCA52012.2021.00047},
  urldate = {2025-08-11},
  copyright = {https://ieeexplore.ieee.org/Xplorehelp/downloads/license-information/IEEE.html},
  isbn = {978-1-6654-3333-4}
}

@inproceedings{haoSupportingAddressTranslation2017,
  title = {Supporting {{Address Translation}} for {{Accelerator-Centric Architectures}}},
  booktitle = {2017 {{IEEE International Symposium}} on {{High Performance Computer Architecture}} ({{HPCA}})},
  author = {Hao, Yuchen and Fang, Zhenman and Reinman, Glenn and Cong, Jason},
  year = 2017,
  month = feb,
  pages = {37--48},
  publisher = {IEEE},
  address = {Austin, TX},
  doi = {10.1109/HPCA.2017.19},
  urldate = {2025-10-09},
  isbn = {978-1-5090-4985-1}
}

@inproceedings{hashemiLearningMemoryAccess2018,
  title = {Learning {{Memory Access Patterns}}},
  booktitle = {Proceedings of the 35th {{International Conference}} on {{Machine Learning}}, {{ICML}} 2018, {{Stockholmsm{\"a}ssan}}, {{Stockholm}}, {{Sweden}}, {{July}} 10-15, 2018},
  author = {Hashemi, Milad and Swersky, Kevin and Smith, Jamie A. and Ayers, Grant and Litz, Heiner and Chang, Jichuan and Kozyrakis, Christos and Ranganathan, Parthasarathy},
  editor = {Dy, Jennifer G. and Krause, Andreas},
  year = 2018,
  month = jul,
  series = {Proceedings of {{Machine Learning Research}}},
  volume = {80},
  pages = {1924--1933},
  publisher = {PMLR}
}

@book{hennessyComputerArchitectureQuantitative2019,
  title = {Computer Architecture: A Quantitative Approach},
  shorttitle = {Computer Architecture},
  author = {Hennessy, John L.},
  year = 2019,
  edition = {Sixth edition},
  publisher = {Morgan Kaufmann Publishers},
  address = {Cambridge, MA},
  isbn = {978-0-12-811905-1},
  lccn = {QA76.9.A73 P377 2019}
}

@article{hennessyNewGoldenAge2019,
  title = {A New Golden Age for Computer Architecture},
  author = {Hennessy, John L. and Patterson, David A.},
  year = 2019,
  month = jan,
  journal = {Communications of the ACM},
  volume = {62},
  number = {2},
  pages = {48--60},
  issn = {0001-0782, 1557-7317},
  doi = {10.1145/3282307},
  urldate = {2025-08-09},
  langid = {english}
}

@misc{hill21stCenturyComputer2016,
  title = {21st {{Century Computer Architecture}}.},
  author = {Hill, Mark D. and Adve, Sarita and Ceze, Luis and Irwin, Mary Jane and Kaeli, David and Martonosi, Margaret and Torrellas, Josep and Wenisch, Thomas F. and Wood, David and Yelick, Katherine},
  year = 2016,
  month = sep,
  doi = {10.48550/ARXIV.1609.06756},
  urldate = {2025-08-09},
  copyright = {arXiv.org perpetual, non-exclusive license}
}

@inproceedings{ibnziadNoFATArchitecturalSupport2021,
  title = {No-{{FAT}}: {{Architectural Support}} for {{Low Overhead Memory Safety Checks}}},
  shorttitle = {No-{{FAT}}},
  booktitle = {2021 {{ACM}}/{{IEEE}} 48th {{Annual International Symposium}} on {{Computer Architecture}} ({{ISCA}})},
  author = {Ibn Ziad, Mohamed Tarek and Arroyo, Miguel A. and Manzhosov, Evgeny and Piersma, Ryan and Sethumadhavan, Simha},
  year = 2021,
  month = jun,
  pages = {916--929},
  publisher = {IEEE},
  address = {Valencia, Spain},
  doi = {10.1109/ISCA52012.2021.00076},
  urldate = {2025-08-09},
  copyright = {https://ieeexplore.ieee.org/Xplorehelp/downloads/license-information/IEEE.html},
  isbn = {978-1-6654-3333-4}
}

@inproceedings{ishikawaDesignObjectOriented1984,
  title = {The Design of an Object Oriented Architecture},
  booktitle = {Proceedings of the 11th Annual International Symposium on {{Computer}} Architecture  - {{ISCA}} '84},
  author = {Ishikawa, Yutaka and Tokoro, Mario},
  year = 1984,
  month = jan,
  pages = {178--187},
  publisher = {ACM Press},
  address = {Not Known},
  doi = {10.1145/800015.808181},
  urldate = {2025-08-10},
  copyright = {https://www.acm.org/publications/policies/copyright\_policy\#Background},
  isbn = {978-0-8186-0538-3},
  langid = {english}
}

@article{jacobVirtualMemoryContemporary1998,
  title = {Virtual Memory in Contemporary Microprocessors},
  author = {Jacob, B. and Mudge, T.},
  year = 1998,
  month = jul,
  journal = {IEEE Micro},
  volume = {18},
  number = {4},
  pages = {60--75},
  issn = {02721732},
  doi = {10.1109/40.710872},
  urldate = {2025-08-18},
  copyright = {https://ieeexplore.ieee.org/Xplorehelp/downloads/license-information/IEEE.html}
}

@article{jagadishBigDataIts2014,
  title = {Big Data and Its Technical Challenges},
  author = {Jagadish, H. V. and Gehrke, Johannes and Labrinidis, Alexandros and Papakonstantinou, Yannis and Patel, Jignesh M. and Ramakrishnan, Raghu and Shahabi, Cyrus},
  year = 2014,
  month = jul,
  journal = {Communications of the ACM},
  volume = {57},
  number = {7},
  pages = {86--94},
  issn = {0001-0782, 1557-7317},
  doi = {10.1145/2611567},
  urldate = {2025-09-10},
  langid = {english}
}

@article{jeroTAGTaggedArchitecture2023,
  title = {{{TAG}}: {{Tagged Architecture Guide}}},
  shorttitle = {{{TAG}}},
  author = {Jero, Samuel and Burow, Nathan and Ward, Bryan and Skowyra, Richard and Khazan, Roger and Shrobe, Howard and Okhravi, Hamed},
  year = 2023,
  month = jul,
  journal = {ACM Computing Surveys},
  volume = {55},
  number = {6},
  pages = {1--34},
  issn = {0360-0300, 1557-7341},
  doi = {10.1145/3533704},
  urldate = {2025-08-10},
  langid = {english}
}

@inproceedings{jiUnderstandingObjectlevelMemory2017,
  title = {Understanding Object-Level Memory Access Patterns across the Spectrum},
  booktitle = {Proceedings of the {{International Conference}} for {{High Performance Computing}}, {{Networking}}, {{Storage}} and {{Analysis}}},
  author = {Ji, Xu and Wang, Chao and {El-Sayed}, Nosayba and Ma, Xiaosong and Kim, Youngjae and Vazhkudai, Sudharshan S. and Xue, Wei and Sanchez, Daniel},
  year = 2017,
  month = nov,
  pages = {1--12},
  publisher = {ACM},
  address = {Denver Colorado},
  doi = {10.1145/3126908.3126917},
  urldate = {2025-08-11},
  isbn = {978-1-4503-5114-0},
  langid = {english}
}

@inproceedings{joannouEfficientTaggedMemory2017,
  title = {Efficient {{Tagged Memory}}},
  booktitle = {2017 {{IEEE International Conference}} on {{Computer Design}} ({{ICCD}})},
  author = {Joannou, Alexandre and Woodruff, Jonathan and Kovacsics, Robert and Moore, Simon W. and Bradbury, Alex and Xia, Hongyan and Watson, Robert N.M. and Chisnall, David and Roe, Michael and Davis, Brooks and Napierala, Edward and Baldwin, John and Gudka, Khilan and Neumann, Peter G. and Mazzinghi, Alfredo and Richardson, Alex and Son, Stacey and Markettos, A. Theodore},
  year = 2017,
  month = nov,
  pages = {641--648},
  publisher = {IEEE},
  address = {Boston, MA},
  doi = {10.1109/ICCD.2017.112},
  urldate = {2025-09-29},
  isbn = {978-1-5386-2254-4}
}

@inproceedings{jungDeepUMTensorMigration2023,
  title = {{{DeepUM}}: {{Tensor Migration}} and {{Prefetching}} in {{Unified Memory}}},
  shorttitle = {{{DeepUM}}},
  booktitle = {Proceedings of the 28th {{ACM International Conference}} on {{Architectural Support}} for {{Programming Languages}} and {{Operating Systems}}, {{Volume}} 2},
  author = {Jung, Jaehoon and Kim, Jinpyo and Lee, Jaejin},
  year = 2023,
  month = jan,
  pages = {207--221},
  publisher = {ACM},
  address = {Vancouver BC Canada},
  doi = {10.1145/3575693.3575736},
  urldate = {2025-10-10},
  isbn = {978-1-4503-9916-6},
  langid = {english}
}

@inproceedings{kanevProfilingWarehousescaleComputer2015,
  title = {Profiling a Warehouse-Scale Computer},
  booktitle = {Proceedings of the 42nd {{Annual International Symposium}} on {{Computer Architecture}}},
  author = {Kanev, Svilen and Darago, Juan Pablo and Hazelwood, Kim and Ranganathan, Parthasarathy and Moseley, Tipp and Wei, Gu-Yeon and Brooks, David},
  year = 2015,
  month = jun,
  pages = {158--169},
  publisher = {ACM},
  address = {Portland Oregon},
  doi = {10.1145/2749469.2750392},
  urldate = {2025-08-09},
  isbn = {978-1-4503-3402-0},
  langid = {english}
}

@inproceedings{karakostasRedundantMemoryMappings2015,
  title = {Redundant Memory Mappings for Fast Access to Large Memories},
  booktitle = {Proceedings of the 42nd {{Annual International Symposium}} on {{Computer Architecture}}},
  author = {Karakostas, Vasileios and Gandhi, Jayneel and Ayar, Furkan and Cristal, Adri{\'a}n and Hill, Mark D. and McKinley, Kathryn S. and Nemirovsky, Mario and Swift, Michael M. and {\"U}nsal, Osman},
  year = 2015,
  month = jun,
  pages = {66--78},
  publisher = {ACM},
  address = {Portland Oregon},
  doi = {10.1145/2749469.2749471},
  urldate = {2025-08-12},
  isbn = {978-1-4503-3402-0},
  langid = {english}
}

@book{kernighanProgrammingLanguage2014,
  title = {The {{C}} Programming Language},
  author = {Kernighan, Brian W. and Ritchie, Dennis M.},
  year = 2014,
  series = {Prentice-{{Hall}} Software Series},
  edition = {2. ed., 52. print},
  publisher = {Prentice-Hall PTR},
  address = {Upper Saddle River, NJ},
  isbn = {978-0-13-110362-7 978-0-13-110370-2},
  langid = {english}
}

@inproceedings{khanSamplingDeadBlock2010,
  title = {Sampling {{Dead Block Prediction}} for {{Last-Level Caches}}},
  booktitle = {2010 43rd {{Annual IEEE}}/{{ACM International Symposium}} on {{Microarchitecture}}},
  author = {Khan, Samira Manabi and Tian, Yingying and Jimenez, Daniel A.},
  year = 2010,
  month = dec,
  pages = {175--186},
  publisher = {IEEE},
  address = {Atlanta, GA, USA},
  doi = {10.1109/MICRO.2010.24},
  urldate = {2025-08-12},
  isbn = {978-1-4244-9071-4}
}

@inproceedings{kimHardwarebasedAlwaysOnHeap2020,
  title = {Hardware-Based {{Always-On Heap Memory Safety}}},
  booktitle = {2020 53rd {{Annual IEEE}}/{{ACM International Symposium}} on {{Microarchitecture}} ({{MICRO}})},
  author = {Kim, Yonghae and Lee, Jaekyu and Kim, Hyesoon},
  year = 2020,
  month = oct,
  pages = {1153--1166},
  publisher = {IEEE},
  address = {Athens, Greece},
  doi = {10.1109/MICRO50266.2020.00095},
  urldate = {2025-08-10},
  copyright = {https://ieeexplore.ieee.org/Xplorehelp/downloads/license-information/IEEE.html},
  isbn = {978-1-7281-7383-2}
}

@inproceedings{kimTypedArchitecturesArchitectural2017,
  title = {Typed {{Architectures}}: {{Architectural Support}} for {{Lightweight Scripting}}},
  shorttitle = {Typed {{Architectures}}},
  booktitle = {Proceedings of the {{Twenty-Second International Conference}} on {{Architectural Support}} for {{Programming Languages}} and {{Operating Systems}}},
  author = {Kim, Channoh and Kim, Jaehyeok and Kim, Sungmin and Kim, Dooyoung and Kim, Namho and Na, Gitae and Oh, Young H. and Cho, Hyeon Gyu and Lee, Jae W.},
  year = 2017,
  month = apr,
  pages = {77--90},
  publisher = {ACM},
  address = {Xi'an China},
  doi = {10.1145/3037697.3037726},
  urldate = {2025-10-10},
  isbn = {978-1-4503-4465-4},
  langid = {english}
}

@article{kocherSpectreAttacksExploiting2020,
  title = {Spectre Attacks: Exploiting Speculative Execution},
  shorttitle = {Spectre Attacks},
  author = {Kocher, Paul and Horn, Jann and Fogh, Anders and Genkin, Daniel and Gruss, Daniel and Haas, Werner and Hamburg, Mike and Lipp, Moritz and Mangard, Stefan and Prescher, Thomas and Schwarz, Michael and Yarom, Yuval},
  year = 2020,
  month = jun,
  journal = {Communications of the ACM},
  volume = {63},
  number = {7},
  pages = {93--101},
  issn = {0001-0782, 1557-7317},
  doi = {10.1145/3399742},
  urldate = {2025-08-28},
  langid = {english}
}

@inproceedings{koldingerArchitectureSupportSingle1992,
  title = {Architecture Support for Single Address Space Operating Systems},
  booktitle = {Proceedings of the Fifth International Conference on {{Architectural}} Support for Programming Languages and Operating Systems},
  author = {Koldinger, Eric J. and Chase, Jeffrey S. and Eggers, Susan J.},
  year = 1992,
  month = sep,
  pages = {175--186},
  publisher = {ACM},
  address = {Boston Massachusetts USA},
  doi = {10.1145/143365.143508},
  urldate = {2025-08-31},
  isbn = {978-0-89791-534-2},
  langid = {english}
}

@article{krishnakumarALEXIAProcessorLightweight2019,
  title = {{{ALEXIA}}: {{A Processor}} with {{Lightweight Extensions}} for {{Memory Safety}}},
  shorttitle = {{{ALEXIA}}},
  author = {Krishnakumar, Gnanambikai and Reddy, Kommuru Alekhya and Rebeiro, Chester},
  year = 2019,
  month = nov,
  journal = {ACM Transactions on Embedded Computing Systems},
  volume = {18},
  number = {6},
  pages = {1--27},
  issn = {1539-9087, 1558-3465},
  doi = {10.1145/3362064},
  urldate = {2025-09-02},
  langid = {english}
}

@inproceedings{kumarMETALCachingMultilevel2024,
  title = {{{METAL}}: {{Caching Multi-level Indexes}} in {{Domain-Specific Architectures}}},
  shorttitle = {{{METAL}}},
  booktitle = {Proceedings of the 29th {{ACM International Conference}} on {{Architectural Support}} for {{Programming Languages}} and {{Operating Systems}}, {{Volume}} 2},
  author = {Kumar, Anagha Molakalmur Anil and Prasanna, Aditya and Balkind, Jonathan and Shriraman, Arrvindh},
  year = 2024,
  month = apr,
  pages = {715--729},
  publisher = {ACM},
  address = {La Jolla CA USA},
  doi = {10.1145/3620665.3640402},
  urldate = {2025-10-09},
  isbn = {979-8-4007-0385-0},
  langid = {english}
}

@inproceedings{kwonLowfatPointersCompact2013,
  title = {Low-Fat Pointers: Compact Encoding and Efficient Gate-Level Implementation of Fat Pointers for Spatial Safety and Capability-Based Security},
  shorttitle = {Low-Fat Pointers},
  booktitle = {Proceedings of the 2013 {{ACM SIGSAC}} Conference on {{Computer}} \& Communications Security - {{CCS}} '13},
  author = {Kwon, Albert and Dhawan, Udit and Smith, Jonathan M. and Knight, Thomas F. and DeHon, Andre},
  year = 2013,
  month = nov,
  pages = {721--732},
  publisher = {ACM Press},
  address = {Berlin, Germany},
  doi = {10.1145/2508859.2516713},
  urldate = {2025-08-10},
  copyright = {https://www.acm.org/publications/policies/copyright\_policy\#Background},
  isbn = {978-1-4503-2477-9},
  langid = {english}
}

@inproceedings{lampsonProtection1971,
  title = {Protection},
  booktitle = {Proceedings of 5th {{Princeton Symposium}} on {{Information Sciences}} and {{Systems}}},
  author = {Lampson, Butler W.},
  year = 1971,
  month = mar,
  pages = {437--443},
  address = {Princeton New Jersey},
  doi = {10.1145/775265.775268},
  urldate = {2025-08-10},
  langid = {english}
}

@inproceedings{leeLetMeInStillEmploying2025,
  title = {Let-{{Me-In}}: ({{Still}}) {{Employing In-pointer Bounds Metadata}} for {{Fine-grained GPU Memory Safety}}},
  shorttitle = {Let-{{Me-In}}},
  booktitle = {2025 {{IEEE International Symposium}} on {{High Performance Computer Architecture}} ({{HPCA}})},
  author = {Lee, Jaewon and Chung, Euijun and Singh, Saurabh and Na, Seonjin and Kim, Yonghae and Lee, Jaekyu and Kim, Hyesoon},
  year = 2025,
  month = mar,
  pages = {1648--1661},
  publisher = {IEEE},
  address = {Las Vegas, NV, USA},
  doi = {10.1109/HPCA61900.2025.00122},
  urldate = {2025-09-02},
  copyright = {https://doi.org/10.15223/policy-029},
  isbn = {979-8-3315-0647-6}
}

@inproceedings{leePIMMMUMemoryManagement2024,
  title = {{{PIM-MMU}}: {{A Memory Management Unit}} for {{Accelerating Data Transfers}} in {{Commercial PIM Systems}}},
  shorttitle = {{{PIM-MMU}}},
  booktitle = {2024 57th {{IEEE}}/{{ACM International Symposium}} on {{Microarchitecture}} ({{MICRO}})},
  author = {Lee, Dongjae and Hyun, Bongjoon and Kim, Taehun and Rhu, Minsoo},
  year = 2024,
  month = nov,
  pages = {627--642},
  publisher = {IEEE},
  address = {Austin, TX, USA},
  doi = {10.1109/MICRO61859.2024.00053},
  urldate = {2025-09-19},
  copyright = {https://doi.org/10.15223/policy-029},
  isbn = {979-8-3503-5057-9}
}

@inproceedings{leeSecuringGPURegionbased2022,
  title = {Securing {{GPU}} via Region-Based Bounds Checking},
  booktitle = {Proceedings of the 49th {{Annual International Symposium}} on {{Computer Architecture}}},
  author = {Lee, Jaewon and Kim, Yonghae and Cao, Jiashen and Kim, Euna and Lee, Jaekyu and Kim, Hyesoon},
  year = 2022,
  month = jun,
  pages = {27--41},
  publisher = {ACM},
  address = {New York New York},
  doi = {10.1145/3470496.3527420},
  urldate = {2025-08-09},
  isbn = {978-1-4503-8610-4},
  langid = {english}
}

@inproceedings{leijenMimallocFreeList2019,
  title = {Mimalloc: {{Free List Sharding}} in {{Action}}},
  booktitle = {Programming {{Languages}} and {{Systems}}},
  author = {Leijen, Daan and Zorn, Benjamin and {de Moura}, Leonardo},
  editor = {Lin, Anthony Widjaja},
  year = 2019,
  month = nov,
  pages = {244--265},
  publisher = {Springer International Publishing},
  address = {Cham},
  isbn = {978-3-030-34175-6},
  langid = {american}
}

@inproceedings{lemayCryptographicCapabilityComputing2021,
  title = {Cryptographic {{Capability Computing}}},
  booktitle = {{{MICRO-54}}: 54th {{Annual IEEE}}/{{ACM International Symposium}} on {{Microarchitecture}}},
  author = {LeMay, Michael and Rakshit, Joydeep and Deutsch, Sergej and Durham, David M. and Ghosh, Santosh and Nori, Anant and Gaur, Jayesh and Weiler, Andrew and Sultana, Salmin and Grewal, Karanvir and Subramoney, Sreenivas},
  year = 2021,
  month = oct,
  pages = {253--267},
  publisher = {ACM},
  address = {Virtual Event Greece},
  doi = {10.1145/3466752.3480076},
  urldate = {2025-09-05},
  isbn = {978-1-4503-8557-2},
  langid = {english}
}

@book{levyCapabilitybasedComputerSystems1984,
  title = {Capability-Based Computer Systems},
  author = {Levy, Henry M.},
  year = 1984,
  publisher = {Digital Press},
  address = {Bedford, Mass},
  isbn = {978-0-932376-22-0},
  lccn = {QA76.9.A73 L48 1984}
}

@article{liaoMovingExascaleZettascale2018,
  title = {Moving from Exascale to Zettascale Computing: Challenges and Techniques},
  shorttitle = {Moving from Exascale to Zettascale Computing},
  author = {Liao, Xiang-ke and Lu, Kai and Yang, Can-qun and Li, Jin-wen and Yuan, Yuan and Lai, Ming-che and Huang, Li-bo and Lu, Ping-jing and Fang, Jian-bin and Ren, Jing and Shen, Jie},
  year = 2018,
  month = oct,
  journal = {Frontiers of Information Technology \& Electronic Engineering},
  volume = {19},
  number = {10},
  pages = {1236--1244},
  issn = {2095-9184, 2095-9230},
  doi = {10.1631/FITEE.1800494},
  urldate = {2025-09-23},
  langid = {english}
}

@inproceedings{liljestrandPACItPointer2019,
  title = {{{PAC}} It up: {{Towards Pointer Integrity}} Using {{ARM Pointer Authentication}}},
  booktitle = {28th {{USENIX Security Symposium}} ({{USENIX Security}} 19)},
  author = {Liljestrand, Hans and Nyman, Thomas and Wang, Kui and Perez, Carlos Chinea and Ekberg, Jan-Erik and Asokan, N.},
  year = 2019,
  month = aug,
  pages = {177--194},
  publisher = {USENIX Association},
  address = {Santa Clara, CA},
  isbn = {978-1-939133-06-9}
}

@inproceedings{limDisaggregatedMemoryExpansion2009,
  title = {Disaggregated Memory for Expansion and Sharing in Blade Servers},
  booktitle = {Proceedings of the 36th Annual International Symposium on {{Computer}} Architecture},
  author = {Lim, Kevin and Chang, Jichuan and Mudge, Trevor and Ranganathan, Parthasarathy and Reinhardt, Steven K. and Wenisch, Thomas F.},
  year = 2009,
  month = jun,
  pages = {267--278},
  publisher = {ACM},
  address = {Austin TX USA},
  doi = {10.1145/1555754.1555789},
  urldate = {2025-08-09},
  isbn = {978-1-60558-526-0},
  langid = {english}
}

@inproceedings{linDrGPUMGuidingMemory2023,
  title = {{{DrGPUM}}: {{Guiding Memory Optimization}} for {{GPU-Accelerated Applications}}},
  shorttitle = {{{DrGPUM}}},
  booktitle = {Proceedings of the 28th {{ACM International Conference}} on {{Architectural Support}} for {{Programming Languages}} and {{Operating Systems}}, {{Volume}} 3},
  author = {Lin, Mao and Zhou, Keren and Su, Pengfei},
  year = 2023,
  month = mar,
  pages = {164--178},
  publisher = {ACM},
  address = {Vancouver BC Canada},
  doi = {10.1145/3582016.3582044},
  urldate = {2025-08-14},
  isbn = {978-1-4503-9918-0},
  langid = {english}
}

@inproceedings{linForestAccessawareGPU2025,
  title = {Forest: {{Access-aware GPU UVM Management}}},
  shorttitle = {Forest},
  booktitle = {Proceedings of the 52nd {{Annual International Symposium}} on {{Computer Architecture}}},
  author = {Lin, Mao and Feng, Yuan and Cox, Guilherme and Jeon, Hyeran},
  year = 2025,
  month = jun,
  pages = {137--152},
  publisher = {ACM},
  address = {Tokyo Japan},
  doi = {10.1145/3695053.3731047},
  urldate = {2025-10-09},
  isbn = {979-8-4007-1261-6},
  langid = {english}
}

@inproceedings{liPondCXLBasedMemory2023,
  title = {Pond: {{CXL-Based Memory Pooling Systems}} for {{Cloud Platforms}}},
  shorttitle = {Pond},
  booktitle = {Proceedings of the 28th {{ACM International Conference}} on {{Architectural Support}} for {{Programming Languages}} and {{Operating Systems}}, {{Volume}} 2},
  author = {Li, Huaicheng and Berger, Daniel S. and Hsu, Lisa and Ernst, Daniel and Zardoshti, Pantea and Novakovic, Stanko and Shah, Monish and Rajadnya, Samir and Lee, Scott and Agarwal, Ishwar and Hill, Mark D. and Fontoura, Marcus and Bianchini, Ricardo},
  year = 2023,
  month = jan,
  pages = {574--587},
  publisher = {ACM},
  address = {Vancouver BC Canada},
  doi = {10.1145/3575693.3578835},
  urldate = {2025-08-09},
  isbn = {978-1-4503-9916-6},
  langid = {english}
}

@inproceedings{liSingleAddressSpaceFaaSJord2025,
  title = {Single-{{Address-Space FaaS}} with {{Jord}}},
  booktitle = {Proceedings of the 52nd {{Annual International Symposium}} on {{Computer Architecture}}},
  author = {Li, Yuanlong and Bhattacharyya, Atri and Kumar, Madhur and Bhattacharjee, Abhishek and Etsion, Yoav and Falsafi, Babak and Kashyap, Sanidhya and Payer, Mathias},
  year = 2025,
  month = jun,
  pages = {694--707},
  publisher = {ACM},
  address = {Tokyo Japan},
  doi = {10.1145/3695053.3731108},
  urldate = {2025-09-23},
  isbn = {979-8-4007-1261-6},
  langid = {english}
}

@inproceedings{marufTPPTransparentPage2023,
  title = {{{TPP}}: {{Transparent Page Placement}} for {{CXL-Enabled Tiered-Memory}}},
  shorttitle = {{{TPP}}},
  booktitle = {Proceedings of the 28th {{ACM International Conference}} on {{Architectural Support}} for {{Programming Languages}} and {{Operating Systems}}, {{Volume}} 3},
  author = {Maruf, Hasan Al and Wang, Hao and Dhanotia, Abhishek and Weiner, Johannes and Agarwal, Niket and Bhattacharya, Pallab and Petersen, Chris and Chowdhury, Mosharaf and Kanaujia, Shobhit and Chauhan, Prakash},
  year = 2023,
  month = mar,
  pages = {742--755},
  publisher = {ACM},
  address = {Vancouver BC Canada},
  doi = {10.1145/3582016.3582063},
  urldate = {2025-08-09},
  isbn = {978-1-4503-9918-0},
  langid = {english}
}

@article{mayerArchitectureBurroughsB50001982,
  title = {The Architecture of the {{Burroughs B5000}}: 20 Years Later and Still Ahead of the Times?},
  shorttitle = {The Architecture of the {{Burroughs B5000}}},
  author = {Mayer, Alastair J. W.},
  year = 1982,
  month = jun,
  journal = {ACM SIGARCH Computer Architecture News},
  volume = {10},
  number = {4},
  pages = {3--10},
  issn = {0163-5964},
  doi = {10.1145/641542.641543},
  urldate = {2025-09-30},
  langid = {english}
}

@inproceedings{menonShaktiTRISCVProcessor2017,
  title = {Shakti-{{T}}: {{A RISC-V Processor}} with {{Light Weight Security Extensions}}},
  shorttitle = {Shakti-{{T}}},
  booktitle = {Proceedings of the {{Hardware}} and {{Architectural Support}} for {{Security}} and {{Privacy}}},
  author = {Menon, Arjun and Murugan, Subadra and Rebeiro, Chester and Gala, Neel and Veezhinathan, Kamakoti},
  year = 2017,
  month = jun,
  pages = {1--8},
  publisher = {ACM},
  address = {Toronto ON Canada},
  doi = {10.1145/3092627.3092629},
  urldate = {2025-09-02},
  isbn = {978-1-4503-5266-6},
  langid = {english}
}

@inproceedings{meswaniHeterogeneousMemoryArchitectures2015,
  title = {Heterogeneous Memory Architectures: {{A HW}}/{{SW}} Approach for Mixing Die-Stacked and off-Package Memories},
  shorttitle = {Heterogeneous Memory Architectures},
  booktitle = {2015 {{IEEE}} 21st {{International Symposium}} on {{High Performance Computer Architecture}} ({{HPCA}})},
  author = {Meswani, Mitesh R. and Blagodurov, Sergey and Roberts, David and Slice, John and Ignatowski, Mike and Loh, Gabriel H.},
  year = 2015,
  month = feb,
  pages = {126--136},
  publisher = {IEEE},
  address = {Burlingame, CA, USA},
  doi = {10.1109/HPCA.2015.7056027},
  urldate = {2025-09-23},
  isbn = {978-1-4799-8930-0}
}

@techreport{millerTrendsChallengesStrategic2019,
  title = {Trends, Challenges, and Strategic Shifts in the Software Vulnerability Mitigation Landscape.},
  author = {Miller, Matt},
  year = 2019,
  month = feb,
  urldate = {2025-08-05},
  langid = {english}
}

@inproceedings{mutluMemoryCentricComputingSolving2025,
  title = {Memory-{{Centric Computing}}: {{Solving Computing}}'s {{Memory Problem}}},
  shorttitle = {Memory-{{Centric Computing}}},
  booktitle = {2025 {{IEEE International Memory Workshop}} ({{IMW}})},
  author = {Mutlu, Onur and Olgun, Ataberk and Y{\"u}ksel, {\.I}smail Emir},
  year = 2025,
  month = may,
  pages = {1--4},
  publisher = {IEEE},
  address = {Monterey, CA, USA},
  doi = {10.1109/IMW61990.2025.11026935},
  urldate = {2025-10-23},
  copyright = {https://doi.org/10.15223/policy-029},
  isbn = {979-8-3503-6298-5}
}

@inproceedings{nagarakatteCETSCompilerEnforced2010,
  title = {{{CETS}}: Compiler Enforced Temporal Safety for {{C}}},
  shorttitle = {{{CETS}}},
  booktitle = {Proceedings of the 2010 International Symposium on {{Memory}} Management},
  author = {Nagarakatte, Santosh and Zhao, Jianzhou and Martin, Milo M.K. and Zdancewic, Steve},
  year = 2010,
  month = jun,
  pages = {31--40},
  publisher = {ACM},
  address = {Toronto Ontario Canada},
  doi = {10.1145/1806651.1806657},
  urldate = {2025-10-17},
  isbn = {978-1-4503-0054-4},
  langid = {english}
}

@inproceedings{nagarakatteWatchdogHardwareSafe2012,
  title = {Watchdog: {{Hardware}} for Safe and Secure Manual Memory Management and Full Memory Safety},
  shorttitle = {Watchdog},
  booktitle = {2012 39th {{Annual International Symposium}} on {{Computer Architecture}} ({{ISCA}})},
  author = {Nagarakatte, Santosh and Martin, Milo M. K. and Zdancewic, Steve},
  year = 2012,
  month = jun,
  pages = {189--200},
  publisher = {IEEE},
  address = {Portland, OR, USA},
  doi = {10.1109/ISCA.2012.6237017},
  urldate = {2025-09-02},
  copyright = {https://doi.org/10.15223/policy-029},
  isbn = {978-1-4673-0475-7 978-1-4673-0476-4}
}

@inproceedings{namFRAMERTaggedpointerCapability2019,
  title = {{{FRAMER}}: A Tagged-Pointer Capability System with Memory Safety Applications},
  shorttitle = {{{FRAMER}}},
  booktitle = {Proceedings of the 35th {{Annual Computer Security Applications Conference}}},
  author = {Nam, Myoung Jin and Akritidis, Periklis and Greaves, David J},
  year = 2019,
  month = dec,
  pages = {612--626},
  publisher = {ACM},
  address = {San Juan Puerto Rico USA},
  doi = {10.1145/3359789.3359799},
  urldate = {2025-08-12},
  isbn = {978-1-4503-7628-0},
  langid = {english}
}

@inproceedings{navarro-torresBertiAccurateLocalDelta2022,
  title = {Berti: An {{Accurate Local-Delta Data Prefetcher}}},
  shorttitle = {Berti},
  booktitle = {2022 55th {{IEEE}}/{{ACM International Symposium}} on {{Microarchitecture}} ({{MICRO}})},
  author = {{Navarro-Torres}, Agustin and Panda, Biswabandan and {Alastruey-Benede}, Jesus and Ibanez, Pablo and {Vinals-Yufera}, Victor and Ros, Alberto},
  year = 2022,
  month = oct,
  pages = {975--991},
  publisher = {IEEE},
  address = {Chicago, IL, USA},
  doi = {10.1109/MICRO56248.2022.00072},
  urldate = {2025-08-12},
  copyright = {https://doi.org/10.15223/policy-029},
  isbn = {978-1-6654-6272-3}
}

@inproceedings{neculaCCuredTypesafeRetrofitting2002,
  title = {{{CCured}}: Type-Safe Retrofitting of Legacy Code},
  shorttitle = {{{CCured}}},
  booktitle = {Proceedings of the 29th {{ACM SIGPLAN-SIGACT}} Symposium on {{Principles}} of Programming Languages},
  author = {Necula, George C. and McPeak, Scott and Weimer, Westley},
  year = 2002,
  month = jan,
  pages = {128--139},
  publisher = {ACM},
  address = {Portland Oregon},
  doi = {10.1145/503272.503286},
  urldate = {2025-10-12},
  isbn = {978-1-58113-450-6},
  langid = {english}
}

@inproceedings{needhamCambridgeCAPComputer1977,
  title = {The {{Cambridge CAP}} Computer and Its Protection System},
  booktitle = {Proceedings of the Sixth Symposium on {{Operating}} Systems Principles  - {{SOSP}} '77},
  author = {Needham, R. M. and Walker, R. D.H.},
  year = 1977,
  month = nov,
  pages = {1--10},
  publisher = {ACM Press},
  address = {West Lafayette, Indiana, United States},
  doi = {10.1145/800214.806541},
  urldate = {2025-08-21},
  copyright = {https://www.acm.org/publications/policies/copyright\_policy\#Background},
  langid = {english}
}

@article{oleksenkoIntelMPXExplained2018,
  title = {Intel {{MPX Explained}}: {{A Cross-layer Analysis}} of the {{Intel MPX System Stack}}},
  shorttitle = {Intel {{MPX Explained}}},
  author = {Oleksenko, Oleksii and Kuvaiskii, Dmitrii and Bhatotia, Pramod and Felber, Pascal and Fetzer, Christof},
  year = 2018,
  month = jun,
  journal = {Proceedings of the ACM on Measurement and Analysis of Computing Systems},
  volume = {2},
  number = {2},
  pages = {1--30},
  issn = {2476-1249},
  doi = {10.1145/3224423},
  urldate = {2025-08-11},
  langid = {english}
}

@inproceedings{pakalapatiBouquetInstructionPointers2020,
  title = {Bouquet of {{Instruction Pointers}}: {{Instruction Pointer Classifier-based Spatial Hardware Prefetching}}},
  shorttitle = {Bouquet of {{Instruction Pointers}}},
  booktitle = {2020 {{ACM}}/{{IEEE}} 47th {{Annual International Symposium}} on {{Computer Architecture}} ({{ISCA}})},
  author = {Pakalapati, Samuel and Panda, Biswabandan},
  year = 2020,
  month = may,
  pages = {118--131},
  publisher = {IEEE},
  address = {Valencia, Spain},
  doi = {10.1109/ISCA45697.2020.00021},
  urldate = {2025-08-11},
  copyright = {https://doi.org/10.15223/policy-029},
  isbn = {978-1-7281-4661-4}
}

@article{pattersonCaseReducedInstruction1980,
  title = {The Case for the Reduced Instruction Set Computer},
  author = {Patterson, David A. and Ditzel, David R.},
  year = 1980,
  month = oct,
  journal = {ACM SIGARCH Computer Architecture News},
  volume = {8},
  number = {6},
  pages = {25--33},
  issn = {0163-5964},
  doi = {10.1145/641914.641917},
  urldate = {2025-09-22},
  langid = {english}
}

@inproceedings{peledSemanticLocalityContextbased2015,
  title = {Semantic Locality and Context-Based Prefetching Using Reinforcement Learning},
  booktitle = {Proceedings of the 42nd {{Annual International Symposium}} on {{Computer Architecture}}},
  author = {Peled, Leeor and Mannor, Shie and Weiser, Uri and Etsion, Yoav},
  year = 2015,
  month = jun,
  pages = {285--297},
  publisher = {ACM},
  address = {Portland Oregon},
  doi = {10.1145/2749469.2749473},
  urldate = {2025-08-09},
  isbn = {978-1-4503-3402-0},
  langid = {english}
}

@inproceedings{pengCapuchinTensorbasedGPU2020,
  title = {Capuchin: {{Tensor-based GPU Memory Management}} for {{Deep Learning}}},
  shorttitle = {Capuchin},
  booktitle = {Proceedings of the {{Twenty-Fifth International Conference}} on {{Architectural Support}} for {{Programming Languages}} and {{Operating Systems}}},
  author = {Peng, Xuan and Shi, Xuanhua and Dai, Hulin and Jin, Hai and Ma, Weiliang and Xiong, Qian and Yang, Fan and Qian, Xuehai},
  year = 2020,
  month = mar,
  pages = {891--905},
  publisher = {ACM},
  address = {Lausanne Switzerland},
  doi = {10.1145/3373376.3378505},
  urldate = {2025-08-11},
  isbn = {978-1-4503-7102-5},
  langid = {english}
}

@article{rattnerObjectbasedComputerArchitecture1980,
  title = {Object-Based Computer Architecture},
  author = {Rattner, Justin and Cox, George},
  year = 1980,
  month = oct,
  journal = {ACM SIGARCH Computer Architecture News},
  volume = {8},
  number = {6},
  pages = {4--11},
  issn = {0163-5964},
  doi = {10.1145/641914.641915},
  urldate = {2025-08-10},
  langid = {english}
}

@article{reedExascaleComputingBig2015,
  title = {Exascale Computing and Big Data},
  author = {Reed, Daniel A. and Dongarra, Jack},
  year = 2015,
  month = jun,
  journal = {Communications of the ACM},
  volume = {58},
  number = {7},
  pages = {56--68},
  issn = {0001-0782, 1557-7317},
  doi = {10.1145/2699414},
  urldate = {2025-08-09},
  langid = {english}
}

@article{saileshwarHeapCheckLowcostHardware2022,
  title = {{{HeapCheck}}: {{Low-cost Hardware Support}} for {{Memory Safety}}},
  shorttitle = {{{HeapCheck}}},
  author = {Saileshwar, Gururaj and Boivie, Rick and Chen, Tong and Segal, Benjamin and Buyuktosunoglu, Alper},
  year = 2022,
  month = mar,
  journal = {ACM Transactions on Architecture and Code Optimization},
  volume = {19},
  number = {1},
  pages = {1--24},
  issn = {1544-3566, 1544-3973},
  doi = {10.1145/3495152},
  urldate = {2025-08-10},
  langid = {english}
}

@article{saltzerProtectionControlInformation1974,
  title = {Protection and the Control of Information Sharing in Multics},
  author = {Saltzer, Jerome H.},
  year = 1974,
  month = jul,
  journal = {Communications of the ACM},
  volume = {17},
  number = {7},
  pages = {388--402},
  issn = {0001-0782, 1557-7317},
  doi = {10.1145/361011.361067},
  urldate = {2025-08-10},
  langid = {english}
}

@article{saltzerProtectionInformationComputer1975,
  title = {The Protection of Information in Computer Systems},
  author = {Saltzer, J.H. and Schroeder, M.D.},
  year = 1975,
  month = sep,
  journal = {Proceedings of the IEEE},
  volume = {63},
  number = {9},
  pages = {1278--1308},
  issn = {0018-9219},
  doi = {10.1109/PROC.1975.9939},
  urldate = {2025-08-10},
  copyright = {https://ieeexplore.ieee.org/Xplorehelp/downloads/license-information/IEEE.html}
}

@inproceedings{sasakiPracticalByteGranularMemory2019,
  title = {Practical {{Byte-Granular Memory Blacklisting}} Using {{Califorms}}},
  booktitle = {Proceedings of the 52nd {{Annual IEEE}}/{{ACM International Symposium}} on {{Microarchitecture}}},
  author = {Sasaki, Hiroshi and Arroyo, Miguel A. and Ziad, M. Tarek Ibn and Bhat, Koustubha and Sinha, Kanad and Sethumadhavan, Simha},
  year = 2019,
  month = oct,
  pages = {558--571},
  publisher = {ACM},
  address = {Columbus OH USA},
  doi = {10.1145/3352460.3358299},
  urldate = {2025-08-11},
  isbn = {978-1-4503-6938-1},
  langid = {english}
}

@inproceedings{serebryanyAddressSanitizerFastAddress2012,
  title = {{{AddressSanitizer}}: {{A Fast Address Sanity Checker}}},
  booktitle = {2012 {{USENIX Annual Technical Conference}} ({{USENIX ATC}} 12)},
  author = {Serebryany, Konstantin and Bruening, Derek and Potapenko, Alexander and Vyukov, Dmitriy},
  year = 2012,
  month = jun,
  pages = {309--318},
  publisher = {USENIX Association},
  address = {Boston, MA},
  isbn = {978-931971-93-5}
}

@inproceedings{sharifiCHEx86ContextSensitiveEnforcement2020,
  title = {{{CHEx86}}: {{Context-Sensitive Enforcement}} of {{Memory Safety}} via {{Microcode-Enabled Capabilities}}},
  shorttitle = {{{CHEx86}}},
  booktitle = {2020 {{ACM}}/{{IEEE}} 47th {{Annual International Symposium}} on {{Computer Architecture}} ({{ISCA}})},
  author = {Sharifi, Rasool and Venkat, Ashish},
  year = 2020,
  month = may,
  pages = {762--775},
  publisher = {IEEE},
  address = {Valencia, Spain},
  doi = {10.1109/ISCA45697.2020.00068},
  urldate = {2025-08-11},
  copyright = {https://doi.org/10.15223/policy-029},
  isbn = {978-1-7281-4661-4}
}

@inproceedings{shiApplyingDeepLearning2019,
  title = {Applying {{Deep Learning}} to the {{Cache Replacement Problem}}},
  booktitle = {Proceedings of the 52nd {{Annual IEEE}}/{{ACM International Symposium}} on {{Microarchitecture}}},
  author = {Shi, Zhan and Huang, Xiangru and Jain, Akanksha and Lin, Calvin},
  year = 2019,
  month = oct,
  pages = {413--425},
  publisher = {ACM},
  address = {Columbus OH USA},
  doi = {10.1145/3352460.3358319},
  urldate = {2025-08-12},
  isbn = {978-1-4503-6938-1},
  langid = {english}
}

@inproceedings{shiHierarchicalNeuralModel2021,
  title = {A Hierarchical Neural Model of Data Prefetching},
  booktitle = {Proceedings of the 26th {{ACM International Conference}} on {{Architectural Support}} for {{Programming Languages}} and {{Operating Systems}}},
  author = {Shi, Zhan and Jain, Akanksha and Swersky, Kevin and Hashemi, Milad and Ranganathan, Parthasarathy and Lin, Calvin},
  year = 2021,
  month = apr,
  pages = {861--873},
  publisher = {ACM},
  address = {Virtual USA},
  doi = {10.1145/3445814.3446752},
  urldate = {2025-08-12},
  isbn = {978-1-4503-8317-2},
  langid = {english}
}

@inproceedings{shullNoMapSpeedingUpJavaScript2019,
  title = {{{NoMap}}: {{Speeding-Up JavaScript Using Hardware Transactional Memory}}},
  shorttitle = {{{NoMap}}},
  booktitle = {2019 {{IEEE International Symposium}} on {{High Performance Computer Architecture}} ({{HPCA}})},
  author = {Shull, Thomas and Choi, Jiho and Garzaran, Maria J. and Torrellas, Josep},
  year = 2019,
  month = feb,
  pages = {412--425},
  publisher = {IEEE},
  address = {Washington, DC, USA},
  doi = {10.1109/HPCA.2019.00054},
  urldate = {2025-08-28},
  copyright = {https://ieeexplore.ieee.org/Xplorehelp/downloads/license-information/IEEE.html},
  isbn = {978-1-7281-1444-6}
}

@inproceedings{sinhaPracticalMemorySafety2018,
  title = {Practical {{Memory Safety}} with {{REST}}},
  booktitle = {2018 {{ACM}}/{{IEEE}} 45th {{Annual International Symposium}} on {{Computer Architecture}} ({{ISCA}})},
  author = {Sinha, Kanad and Sethumadhavan, Simha},
  year = 2018,
  month = jun,
  pages = {600--611},
  publisher = {IEEE},
  address = {Los Angeles, CA},
  doi = {10.1109/ISCA.2018.00056},
  urldate = {2025-10-11},
  isbn = {978-1-5386-5984-7}
}

@inproceedings{songHDFIHardwareAssistedDataFlow2016,
  title = {{{HDFI}}: {{Hardware-Assisted Data-Flow Isolation}}},
  shorttitle = {{{HDFI}}},
  booktitle = {2016 {{IEEE Symposium}} on {{Security}} and {{Privacy}} ({{SP}})},
  author = {Song, Chengyu and Moon, Hyungon and Alam, Monjur and Yun, Insu and Lee, Byoungyoung and Kim, Taesoo and Lee, Wenke and Paek, Yunheung},
  year = 2016,
  month = may,
  pages = {1--17},
  publisher = {IEEE},
  address = {San Jose, CA},
  doi = {10.1109/SP.2016.9},
  urldate = {2025-10-12},
  isbn = {978-1-5090-0824-7}
}

@inproceedings{songSoKSanitizingSecurity2019,
  title = {{{SoK}}: {{Sanitizing}} for {{Security}}},
  shorttitle = {{{SoK}}},
  booktitle = {2019 {{IEEE Symposium}} on {{Security}} and {{Privacy}} ({{SP}})},
  author = {Song, Dokyung and Lettner, Julian and Rajasekaran, Prabhu and Na, Yeoul and Volckaert, Stijn and Larsen, Per and Franz, Michael},
  year = 2019,
  month = may,
  pages = {1275--1295},
  publisher = {IEEE},
  address = {San Francisco, CA, USA},
  doi = {10.1109/SP.2019.00010},
  urldate = {2025-08-09},
  copyright = {https://ieeexplore.ieee.org/Xplorehelp/downloads/license-information/IEEE.html},
  isbn = {978-1-5386-6660-9}
}

@inproceedings{su11InnovationNext2023,
  title = {1.1 {{Innovation For}} the {{Next Decade}} of {{Compute Efficiency}}},
  booktitle = {2023 {{IEEE International Solid-State Circuits Conference}} ({{ISSCC}})},
  author = {Su, Lisa and Naffziger, Sam},
  year = 2023,
  month = feb,
  pages = {8--12},
  issn = {2376-8606},
  doi = {10.1109/ISSCC42615.2023.10067810},
  urldate = {2025-09-23}
}

@inproceedings{suhSecureProgramExecution2004,
  title = {Secure Program Execution via Dynamic Information Flow Tracking},
  booktitle = {Proceedings of the 11th International Conference on {{Architectural}} Support for Programming Languages and Operating Systems},
  author = {Suh, G. Edward and Lee, Jae W. and Zhang, David and Devadas, Srinivas},
  year = 2004,
  month = oct,
  pages = {85--96},
  publisher = {ACM},
  address = {Boston MA USA},
  doi = {10.1145/1024393.1024404},
  urldate = {2025-08-10},
  isbn = {978-1-58113-804-7},
  langid = {english}
}

@inproceedings{sullivanImplicitMemoryTagging2023,
  title = {Implicit {{Memory Tagging}}: {{No-Overhead Memory Safety Using Alias-Free Tagged ECC}}},
  shorttitle = {Implicit {{Memory Tagging}}},
  booktitle = {Proceedings of the 50th {{Annual International Symposium}} on {{Computer Architecture}}},
  author = {Sullivan, Michael B. and Ziad, Mohamed Tarek Ibn and Jaleel, Aamer and Keckler, Stephen W.},
  year = 2023,
  month = jun,
  pages = {1--13},
  publisher = {ACM},
  address = {Orlando FL USA},
  doi = {10.1145/3579371.3589102},
  urldate = {2025-09-05},
  isbn = {979-8-4007-0095-8},
  langid = {english}
}

@inproceedings{szekeresSoKEternalWar2013,
  title = {{{SoK}}: {{Eternal War}} in {{Memory}}},
  shorttitle = {{{SoK}}},
  booktitle = {2013 {{IEEE Symposium}} on {{Security}} and {{Privacy}}},
  author = {Szekeres, L. and Payer, M. and {Tao Wei} and Song, Dawn},
  year = 2013,
  month = may,
  pages = {48--62},
  publisher = {IEEE},
  address = {Berkeley, CA},
  doi = {10.1109/SP.2013.13},
  urldate = {2025-08-09},
  isbn = {978-0-7695-4977-4 978-1-4673-6166-8}
}

@inproceedings{thorntonParallelOperationControl1964,
  title = {Parallel Operation in the Control Data 6600},
  booktitle = {Proceedings of the {{October}} 27-29, 1964, Fall Joint Computer Conference, Part {{II}}: Very High Speed Computer Systems on {{XX}} - {{AFIPS}} '64 ({{Fall}}, Part {{II}})},
  author = {Thornton, James E.},
  year = 1964,
  month = oct,
  pages = {33},
  publisher = {ACM Press},
  address = {San Francisco, California},
  doi = {10.1145/1464039.1464045},
  urldate = {2025-08-10},
  copyright = {https://www.acm.org/publications/policies/copyright\_policy\#Background},
  langid = {english}
}

@inproceedings{tiwariSmallCacheLarge2008,
  title = {A Small Cache of Large Ranges: {{Hardware}} Methods for Efficiently Searching, Storing, and Updating Big Dataflow Tags},
  shorttitle = {A Small Cache of Large Ranges},
  booktitle = {2008 41st {{IEEE}}/{{ACM International Symposium}} on {{Microarchitecture}}},
  author = {Tiwari, Mohit and Agrawal, Banit and Mysore, Shashidhar and Valamehr, Jonathan and Sherwood, Timothy},
  year = 2008,
  month = nov,
  pages = {94--105},
  publisher = {IEEE},
  address = {Como, Italy},
  doi = {10.1109/MICRO.2008.4771782},
  urldate = {2025-08-10},
  isbn = {978-1-4244-2836-6}
}

@inproceedings{tsaiCompressObjectsNot2019,
  title = {Compress {{Objects}}, {{Not Cache Lines}}: {{An Object-Based Compressed Memory Hierarchy}}},
  shorttitle = {Compress {{Objects}}, {{Not Cache Lines}}},
  booktitle = {Proceedings of the {{Twenty-Fourth International Conference}} on {{Architectural Support}} for {{Programming Languages}} and {{Operating Systems}}},
  author = {Tsai, Po-An and Sanchez, Daniel},
  year = 2019,
  month = apr,
  pages = {229--242},
  publisher = {ACM},
  address = {Providence RI USA},
  doi = {10.1145/3297858.3304006},
  urldate = {2025-08-11},
  isbn = {978-1-4503-6240-5},
  langid = {english}
}

@inproceedings{tsaiRethinkingMemoryHierarchy2018,
  title = {Rethinking the {{Memory Hierarchy}} for {{Modern Languages}}},
  booktitle = {2018 51st {{Annual IEEE}}/{{ACM International Symposium}} on {{Microarchitecture}} ({{MICRO}})},
  author = {Tsai, Po-An and Gan, Yee Ling and Sanchez, Daniel},
  year = 2018,
  month = oct,
  pages = {203--216},
  publisher = {IEEE},
  address = {Fukuoka},
  doi = {10.1109/MICRO.2018.00025},
  urldate = {2025-08-09},
  isbn = {978-1-5386-6240-3}
}

@inproceedings{unterguggenbergerCryptographicallyEnforcedMemory2023,
  title = {Cryptographically {{Enforced Memory Safety}}},
  booktitle = {Proceedings of the 2023 {{ACM SIGSAC Conference}} on {{Computer}} and {{Communications Security}}},
  author = {Unterguggenberger, Martin and Schrammel, David and Lamster, Lukas and Nasahl, Pascal and Mangard, Stefan},
  year = 2023,
  month = nov,
  pages = {889--903},
  publisher = {ACM},
  address = {Copenhagen Denmark},
  doi = {10.1145/3576915.3623138},
  urldate = {2025-08-09},
  isbn = {979-8-4007-0050-7},
  langid = {english}
}

@inproceedings{vijaykumarCaseRicherCrossLayer2018,
  title = {A {{Case}} for {{Richer Cross-Layer Abstractions}}: {{Bridging}} the {{Semantic Gap}} with {{Expressive Memory}}},
  shorttitle = {A {{Case}} for {{Richer Cross-Layer Abstractions}}},
  booktitle = {2018 {{ACM}}/{{IEEE}} 45th {{Annual International Symposium}} on {{Computer Architecture}} ({{ISCA}})},
  author = {Vijaykumar, Nandita and Jain, Abhilasha and Majumdar, Diptesh and Hsieh, Kevin and Pekhimenko, Gennady and Ebrahimi, Eiman and Hajinazar, Nastaran and Gibbons, Phillip B. and Mutlu, Onur},
  year = 2018,
  month = jun,
  pages = {207--220},
  publisher = {IEEE},
  address = {Los Angeles, CA},
  doi = {10.1109/ISCA.2018.00027},
  urldate = {2025-08-11},
  isbn = {978-1-5386-5984-7}
}

@inproceedings{vijaykumarLocalityDescriptorHolistic2018,
  title = {The {{Locality Descriptor}}: {{A Holistic Cross-Layer Abstraction}} to {{Express Data Locality In GPUs}}},
  shorttitle = {The {{Locality Descriptor}}},
  booktitle = {2018 {{ACM}}/{{IEEE}} 45th {{Annual International Symposium}} on {{Computer Architecture}} ({{ISCA}})},
  author = {Vijaykumar, Nandita and Ebrahimi, Eiman and Hsieh, Kevin and Gibbons, Phillip B. and Mutlu, Onur},
  year = 2018,
  month = jun,
  pages = {829--842},
  publisher = {IEEE},
  address = {Los Angeles, CA},
  doi = {10.1109/ISCA.2018.00074},
  urldate = {2025-08-11},
  isbn = {978-1-5386-5984-7}
}

@article{vijaykumarMetaSysPracticalOpensource2022,
  title = {{{MetaSys}}: {{A Practical Open-source Metadata Management System}} to {{Implement}} and {{Evaluate Cross-layer Optimizations}}},
  shorttitle = {{{MetaSys}}},
  author = {Vijaykumar, Nandita and Olgun, Ataberk and Kanellopoulos, Konstantinos and Bostanci, F. Nisa and Hassan, Hasan and Lotfi, Mehrshad and Gibbons, Phillip B. and Mutlu, Onur},
  year = 2022,
  month = jun,
  journal = {ACM Transactions on Architecture and Code Optimization},
  volume = {19},
  number = {2},
  pages = {1--29},
  issn = {1544-3566, 1544-3973},
  doi = {10.1145/3505250},
  urldate = {2025-08-11},
  langid = {english}
}

@inproceedings{volosMnemosyneLightweightPersistent2011,
  title = {Mnemosyne: Lightweight Persistent Memory},
  shorttitle = {Mnemosyne},
  booktitle = {Proceedings of the Sixteenth International Conference on {{Architectural}} Support for Programming Languages and Operating Systems},
  author = {Volos, Haris and Tack, Andres Jaan and Swift, Michael M.},
  year = 2011,
  month = mar,
  pages = {91--104},
  publisher = {ACM},
  address = {Newport Beach California USA},
  doi = {10.1145/1950365.1950379},
  urldate = {2025-10-27},
  isbn = {978-1-4503-0266-1},
  langid = {english}
}

@inproceedings{wangHardwareSupportedPersistent2017,
  title = {Hardware Supported Persistent Object Address Translation},
  booktitle = {Proceedings of the 50th {{Annual IEEE}}/{{ACM International Symposium}} on {{Microarchitecture}}},
  author = {Wang, Tiancong and Sambasivam, Sakthikumaran and Solihin, Yan and Tuck, James},
  year = 2017,
  month = oct,
  pages = {800--812},
  publisher = {ACM},
  address = {Cambridge Massachusetts},
  doi = {10.1145/3123939.3123981},
  urldate = {2025-08-11},
  isbn = {978-1-4503-4952-9},
  langid = {english}
}

@inproceedings{wangMementoArchitecturalSupport2023,
  title = {Memento: {{Architectural Support}} for {{Ephemeral Memory Management}} in {{Serverless Environments}}},
  shorttitle = {Memento},
  booktitle = {56th {{Annual IEEE}}/{{ACM International Symposium}} on {{Microarchitecture}}},
  author = {Wang, Ziqi and Zhao, Kaiyang and Li, Pei and Jacob, Andrew and Kozuch, Michael and Mowry, Todd and Skarlatos, Dimitrios},
  year = 2023,
  month = oct,
  pages = {122--136},
  publisher = {ACM},
  address = {Toronto ON Canada},
  doi = {10.1145/3613424.3623795},
  urldate = {2025-08-11},
  isbn = {979-8-4007-0329-4},
  langid = {english}
}

@inproceedings{wangOASISObjectAwarePage2025,
  title = {{{OASIS}}: {{Object-Aware Page Management}} for {{Multi-GPU Systems}}},
  shorttitle = {{{OASIS}}},
  booktitle = {2025 {{IEEE International Symposium}} on {{High Performance Computer Architecture}} ({{HPCA}})},
  author = {Wang, Yueqi and Li, Bingyao and Ziad, Mohamed Tarek Ibn and Eeckhout, Lieven and Yang, Jun and Jaleel, Aamer and Tang, Xulong},
  year = 2025,
  month = mar,
  pages = {1678--1692},
  publisher = {IEEE},
  address = {Las Vegas, NV, USA},
  doi = {10.1109/HPCA61900.2025.00124},
  urldate = {2025-08-11},
  copyright = {https://doi.org/10.15223/policy-029},
  isbn = {979-8-3315-0647-6}
}

@inproceedings{watsonCHERIHybridCapabilitySystem2015,
  title = {{{CHERI}}: {{A Hybrid Capability-System Architecture}} for {{Scalable Software Compartmentalization}}},
  shorttitle = {{{CHERI}}},
  booktitle = {2015 {{IEEE Symposium}} on {{Security}} and {{Privacy}}},
  author = {Watson, Robert N.M. and Woodruff, Jonathan and Neumann, Peter G. and Moore, Simon W. and Anderson, Jonathan and Chisnall, David and Dave, Nirav and Davis, Brooks and Gudka, Khilan and Laurie, Ben and Murdoch, Steven J. and Norton, Robert and Roe, Michael and Son, Stacey and Vadera, Munraj},
  year = 2015,
  month = may,
  pages = {20--37},
  issn = {2375-1207},
  doi = {10.1109/SP.2015.9},
  urldate = {2025-09-22}
}

@inproceedings{welchInvestigationDescriptorOriented1976,
  title = {An Investigation of Descriptor Oriented Architecture},
  booktitle = {Proceedings of the 3rd Annual Symposium on {{Computer}} Architecture  - {{ISCA}} '76},
  author = {Welch, Terry A.},
  year = 1976,
  pages = {141--146},
  publisher = {ACM Press},
  address = {Not Known},
  doi = {10.1145/800110.803571},
  urldate = {2025-09-22},
  copyright = {https://www.acm.org/publications/policies/copyright\_policy\#Background},
  langid = {english}
}

@inproceedings{witchelMondrianMemoryProtection2002,
  title = {Mondrian Memory Protection},
  booktitle = {Proceedings of the 10th International Conference on {{Architectural}} Support for Programming Languages and Operating Systems},
  author = {Witchel, Emmett and Cates, Josh and Asanovi{\'c}, Krste},
  year = 2002,
  month = oct,
  pages = {304--316},
  publisher = {ACM},
  address = {San Jose California},
  doi = {10.1145/605397.605429},
  urldate = {2025-08-10},
  isbn = {978-1-58113-574-9},
  langid = {english}
}

@inproceedings{woodruffCHERICapabilityModel2014,
  title = {The {{CHERI}} Capability Model: {{Revisiting RISC}} in an Age of Risk},
  shorttitle = {The {{CHERI}} Capability Model},
  booktitle = {2014 {{ACM}}/{{IEEE}} 41st {{International Symposium}} on {{Computer Architecture}} ({{ISCA}})},
  author = {Woodruff, Jonathan and Watson, Robert N. M. and Chisnall, David and Moore, Simon W. and Anderson, Jonathan and Davis, Brooks and Laurie, Ben and Neumann, Peter G. and Norton, Robert and Roe, Michael},
  year = 2014,
  month = jun,
  pages = {457--468},
  publisher = {IEEE},
  address = {Minneapolis, MN, USA},
  doi = {10.1109/ISCA.2014.6853201},
  urldate = {2025-08-20},
  isbn = {978-1-4799-4394-4 978-1-4799-4396-8}
}

@article{woodruffCHERIConcentratePractical2019,
  title = {{{CHERI Concentrate}}: {{Practical Compressed Capabilities}}},
  shorttitle = {{{CHERI Concentrate}}},
  author = {Woodruff, Jonathan and Joannou, Alexandre and Xia, Hongyan and Fox, Anthony and Norton, Robert M. and Chisnall, David and Davis, Brooks and Gudka, Khilan and Filardo, Nathaniel W. and Markettos, A. Theodore and Roe, Michael and Neumann, Peter G. and Watson, Robert N. M. and Moore, Simon W.},
  year = 2019,
  month = oct,
  journal = {IEEE Transactions on Computers},
  volume = {68},
  number = {10},
  pages = {1455--1469},
  issn = {0018-9340, 1557-9956, 2326-3814},
  doi = {10.1109/TC.2019.2914037},
  urldate = {2025-08-11},
  copyright = {https://ieeexplore.ieee.org/Xplorehelp/downloads/license-information/IEEE.html}
}

@inproceedings{wuSHiPSignaturebasedHit2011,
  title = {{{SHiP}}: Signature-Based Hit Predictor for High Performance Caching},
  shorttitle = {{{SHiP}}},
  booktitle = {Proceedings of the 44th {{Annual IEEE}}/{{ACM International Symposium}} on {{Microarchitecture}}},
  author = {Wu, Carole-Jean and Jaleel, Aamer and Hasenplaugh, Will and Martonosi, Margaret and Steely, Simon C. and Emer, Joel},
  year = 2011,
  month = dec,
  pages = {430--441},
  publisher = {ACM},
  address = {Porto Alegre Brazil},
  doi = {10.1145/2155620.2155671},
  urldate = {2025-08-11},
  isbn = {978-1-4503-1053-6},
  langid = {english}
}

@inproceedings{xiaCHERIvokeCharacterisingPointer2019,
  title = {{{CHERIvoke}}: {{Characterising Pointer Revocation}} Using {{CHERI Capabilities}} for {{Temporal Memory Safety}}},
  shorttitle = {{{CHERIvoke}}},
  booktitle = {Proceedings of the 52nd {{Annual IEEE}}/{{ACM International Symposium}} on {{Microarchitecture}}},
  author = {Xia, Hongyan and Woodruff, Jonathan and Ainsworth, Sam and Filardo, Nathaniel W. and Roe, Michael and Richardson, Alexander and Rugg, Peter and Neumann, Peter G. and Moore, Simon W. and Watson, Robert N. M. and Jones, Timothy M.},
  year = 2019,
  month = oct,
  pages = {545--557},
  publisher = {ACM},
  address = {Columbus OH USA},
  doi = {10.1145/3352460.3358288},
  urldate = {2025-09-02},
  isbn = {978-1-4503-6938-1},
  langid = {english}
}

@inproceedings{xuInfatPointerHardwareassisted2021,
  title = {In-Fat Pointer: Hardware-Assisted Tagged-Pointer Spatial Memory Safety Defense with Subobject Granularity Protection},
  shorttitle = {In-Fat Pointer},
  booktitle = {Proceedings of the 26th {{ACM International Conference}} on {{Architectural Support}} for {{Programming Languages}} and {{Operating Systems}}},
  author = {Xu, Shengjie and Huang, Wei and Lie, David},
  year = 2021,
  month = apr,
  pages = {224--240},
  publisher = {ACM},
  address = {Virtual USA},
  doi = {10.1145/3445814.3446761},
  urldate = {2025-08-10},
  isbn = {978-1-4503-8317-2},
  langid = {english}
}

@inproceedings{zhangBOGOBuySpatial2019,
  title = {{{BOGO}}: {{Buy Spatial Memory Safety}}, {{Get Temporal Memory Safety}} ({{Almost}}) {{Free}}},
  shorttitle = {{{BOGO}}},
  booktitle = {Proceedings of the {{Twenty-Fourth International Conference}} on {{Architectural Support}} for {{Programming Languages}} and {{Operating Systems}}},
  author = {Zhang, Tong and Lee, Dongyoon and Jung, Changhee},
  year = 2019,
  month = apr,
  pages = {631--644},
  publisher = {ACM},
  address = {Providence RI USA},
  doi = {10.1145/3297858.3304017},
  urldate = {2025-08-13},
  isbn = {978-1-4503-6240-5},
  langid = {english}
}

@inproceedings{zhangDirectMemoryTranslation2024,
  title = {Direct {{Memory Translation}} for {{Virtualized Clouds}}},
  booktitle = {Proceedings of the 29th {{ACM International Conference}} on {{Architectural Support}} for {{Programming Languages}} and {{Operating Systems}}, {{Volume}} 2},
  author = {Zhang, Jiyuan and Jia, Weiwei and Chai, Siyuan and Liu, Peizhe and Kim, Jongyul and Xu, Tianyin},
  year = 2024,
  month = apr,
  pages = {287--304},
  publisher = {ACM},
  address = {La Jolla CA USA},
  doi = {10.1145/3620665.3640358},
  urldate = {2025-08-11},
  isbn = {979-8-4007-0385-0},
  langid = {english}
}

@inproceedings{zhangG10EnablingEfficient2023,
  title = {G10: {{Enabling An Efficient Unified GPU Memory}} and {{Storage Architecture}} with {{Smart Tensor Migrations}}},
  shorttitle = {G10},
  booktitle = {56th {{Annual IEEE}}/{{ACM International Symposium}} on {{Microarchitecture}}},
  author = {Zhang, Haoyang and Zhou, Yirui and Xue, Yuqi and Liu, Yiqi and Huang, Jian},
  year = 2023,
  month = oct,
  pages = {395--410},
  publisher = {ACM},
  address = {Toronto ON Canada},
  doi = {10.1145/3613424.3614309},
  urldate = {2025-08-12},
  isbn = {979-8-4007-0329-4},
  langid = {english}
}

@inproceedings{zhangUnderstandingRuntimePerformance2022,
  title = {Towards {{Understanding}} the {{Runtime Performance}} of {{Rust}}},
  booktitle = {Proceedings of the 37th {{IEEE}}/{{ACM International Conference}} on {{Automated Software Engineering}}},
  author = {Zhang, Yuchen and Zhang, Yunhang and Portokalidis, Georgios and Xu, Jun},
  year = 2022,
  month = oct,
  pages = {1--6},
  publisher = {ACM},
  address = {Rochester MI USA},
  doi = {10.1145/3551349.3559494},
  urldate = {2025-08-09},
  isbn = {978-1-4503-9475-8},
  langid = {english}
}

@inproceedings{zhaoContiguitasPursuitPhysical2023,
  title = {Contiguitas: {{The Pursuit}} of {{Physical Memory Contiguity}} in {{Datacenters}}},
  shorttitle = {Contiguitas},
  booktitle = {Proceedings of the 50th {{Annual International Symposium}} on {{Computer Architecture}}},
  author = {Zhao, Kaiyang and Xue, Kaiwen and Wang, Ziqi and Schatzberg, Dan and Yang, Leon and Manousis, Antonis and Weiner, Johannes and Van Riel, Rik and Sharma, Bikash and Tang, Chunqiang and Skarlatos, Dimitrios},
  year = 2023,
  month = jun,
  pages = {1--15},
  publisher = {ACM},
  address = {Orlando FL USA},
  doi = {10.1145/3579371.3589079},
  urldate = {2025-09-22},
  isbn = {979-8-4007-0095-8},
  langid = {english}
}

@inproceedings{zhouCharacterizingMemoryAllocator2024,
  title = {Characterizing a {{Memory Allocator}} at {{Warehouse Scale}}},
  booktitle = {Proceedings of the 29th {{ACM International Conference}} on {{Architectural Support}} for {{Programming Languages}} and {{Operating Systems}}, {{Volume}} 3},
  author = {Zhou, Zhuangzhuang and Gogte, Vaibhav and Vaish, Nilay and Kennelly, Chris and Xia, Patrick and Kanev, Svilen and Moseley, Tipp and Delimitrou, Christina and Ranganathan, Parthasarathy},
  year = 2024,
  month = apr,
  pages = {192--206},
  publisher = {ACM},
  address = {La Jolla CA USA},
  doi = {10.1145/3620666.3651350},
  urldate = {2025-08-11},
  isbn = {979-8-4007-0386-7},
  langid = {english}
}

\end{document}